\documentclass[journal]{IEEEtran}
\usepackage{amssymb}
\usepackage[cmex10]{amsmath}
\usepackage{stfloats}
\usepackage{graphicx}
\usepackage{subfigure}
\usepackage{tabularx}
\usepackage{epsfig,epsf,color,balance,cite}
\usepackage{verbatim}
\usepackage{url}
\usepackage{bm}

\newtheorem{theorem}{Theorem}
\newtheorem{lemma}{Lemma}

\usepackage{algorithm}
\usepackage{algpseudocode}
\usepackage{amsmath}

\begin{document}

\title{\textcolor[rgb]{0.00,0.00,0.00}{Robust Transmission Design for Multi-Cell D2D Underlaid Cellular Networks}}

\author{
\IEEEauthorblockN{Hao Xu\IEEEauthorrefmark{0},
                  Gordon L. St{\"u}ber, \emph{Fellow, IEEE}\IEEEauthorrefmark{0},
                  Wei Xu, \emph{Senior Member, IEEE}\IEEEauthorrefmark{0},
                  Cunhua Pan\IEEEauthorrefmark{0},
                  Jianfeng Shi\IEEEauthorrefmark{0},
                  Zhaohui Yang\IEEEauthorrefmark{0},
 and
                  Ming Chen\IEEEauthorrefmark{0}
                  }
\thanks{Copyright (c) 2015 IEEE. Personal use of this material is permitted. However, permission to use this material for any other purposes must be obtained from the IEEE by sending a request to pubs-permissions@ieee.org. This work was in part supported by the NSFC (Nos. 61372106, 61471114, \& 61221002), NSTMP under 2016ZX03001016-003, the Six Talent Peaks project in Jiangsu Province under GDZB-005, Science and Technology Project of Guangdong Province under Grant 2014B010119001, the Scholarship from the China Scholarship Council (No. 201606090039), Program Sponsored for Scientific Innovation Research of College Graduate in Jiangsu Province under Grant KYLX16\_0221, and the Scientific Research Foundation of Graduate School of Southeast University under Grant YBJJ1651. \emph{(Corresponding author: Hao Xu; Wei Xu.)}}

\thanks{H. Xu, W. Xu, J. Shi, Z. Yang and M. Chen are with the National Mobile Communications Research Laboratory, Southeast University, Nanjing 210096, China (e-mail: xuhao2013@seu.edu.cn; wxu@seu.edu.cn; shijianfeng@seu.edu.cn; yangzhaohui@seu.edu.cn; chenming@seu.edu.cn).

G. L. St{\"u}ber is with the Faculty of Electrical and Computer Engineering, Georgia Institute of Technology, Atlanta, GA 30332 USA (e-mail: stuber@ece.gatech.edu).

C. Pan is with School of Electronic Engineering and Computer Science, Queen Mary University of London, London E1 4NS, U.K. (e-mail: c.pan@qmul.ac.uk).
}}

\maketitle

\begin{abstract}
This paper investigates the robust transmission design (RTD) of a multi-cell device-to-device (D2D) underlaid cellular network with imperfect channel state information (CSI). The bounded model is adopted to characterize the CSI impairment and the aim is to maximize the worst-case sum rate of the system. To protect cellular communications, it is assumed that the interference from all D2D transmitters to each base station (BS) is power-limited. It is first shown that the worst-case signal-to-interference-plus-noise ratio (SINR) of each D2D link can be obtained directly, while that of cellular links cannot be similarly found since the channel estimation error vectors of cellular links are coupled in the SINR expressions. To solve the nonconvex problem, the objective function of the original problem is replaced with its lower bound, and the resulted problem is decomposed into multiple semidefinite programming (SDP) subproblems which are convex and have computationally efficient solutions. An iterative RTD algorithm is then proposed to obtain a suboptimal solution. Simulation results show that D2D communication can significantly increase the performance of the conventional cellular systems while causing tolerable interference to cellular users. In addition, the proposed RTD algorithm outperforms the conventional non-robust transmission design greatly in terms of network spectral efficiency.

\end{abstract}


\IEEEpeerreviewmaketitle

\section{Introduction}
\label{Introduction}
With the daily increasing of wireless communication demand, the problem of spectrum insufficiency has become a major factor limiting the wireless system performance \cite{force2002report, index2016global}. Device-to-device (D2D) communication is a promising method for enhancing the spectral efficiency (SE) of traditional cellular systems and has drawn great attention recently \cite{liu2015device,asadi2014survey, lin2015spectral, he2017se}. Different from the conventional cellular communication where all traffic is routed via base stations (BSs), D2D communication allows two closely located users to communicate directly and, thus, has distinct advantages such as high SE, short packet delay, low energy consumption and increased safety \cite{lin2014overview, doumi2013lte}. In a D2D underlaid cellular network, D2D users (DUs) reuse the resource blocks (RBs) of cellular users (CUs) for communication, leading to cochannel interference. Therefore, efficient resource allocation and power control algorithms play an important role in reaping the potential benefits of D2D communication.

Thus far, there have been a flurry of works studying interference mitigation and sum SE maximization problems of D2D underlaid systems \cite{wei2013device, zhu2014downlink, xu2016channel, hoang2014resource, feng2013device, zhao2015resource, 7448661}. Reference \cite{wei2013device} considered a multiple-input multiple-output (MIMO) D2D underlaid system and aimed to maximize the sum SE of DUs by optimizing the precoding matrix. However, the quality-of-service (QoS) of CUs was not guaranteed. Both \cite{zhu2014downlink} and \cite{xu2016channel} considered resource allocation and power control problems to maximize the sum SE of DUs with all CUs protected by the QoS constraints. It was shown in \cite{zhu2014downlink} and \cite{xu2016channel} that the minimum QoS constraints for cellular links always hold with equality under the optimal conditions. Since CUs usually possess higher priorities compared to DUs, this can be unfair for CUs. In \cite{hoang2014resource, feng2013device, zhao2015resource, 7448661}, the problem of maximizing the sum rate of both cellular and D2D links was studied, and the minimum QoS constraints of CUs were imposed to protect cellular communication.

References \cite{wei2013device, zhu2014downlink, xu2016channel, hoang2014resource, feng2013device, zhao2015resource, 7448661} all considered a simple single-cell scenario, while ignoring the cumulated interference from neighbor cells. According to our survey, the performance of multi-cell D2D underlaid systems has less well been studied \cite{feng2014tractable, della2016potential, lin2015interplay}. In \cite{feng2014tractable}, the authors established a tractable model for multi-cell D2D underlaid cellular networks and adopted Exclusion Regions around the BSs to mitigate cochannel interference. In \cite{della2016potential}, the subcarrier allocation problem for a multi-cell D2D underlaid system was characterized as a potential game, and an iterative algorithm was proposed to obtain a Nash equilibrium. Reference \cite{lin2015interplay} considered a multi-cell D2D underlaid massive MIMO system, and investigated the SE of cellular as well as D2D links under both perfect and imperfect channel state information (CSI).

Most of the aforementioned works assumed that perfect CSI was available for system performance analysis and optimization \cite{wei2013device, zhu2014downlink, xu2016channel, hoang2014resource, feng2013device, zhao2015resource, 7448661, feng2014tractable, della2016potential}. However, in practice, it is difficult to obtain perfect CSI of all links due to channel estimation errors and quantization errors, especially in multi-cell systems. Therefore, enhancing the robustness of network performance under partial or imperfect CSI has become an important issue \cite{shenouda2008design, zhang2008statistically, vucic2009robust, tajer2011robust, hanif2013efficient, xu2015robust}. According to \cite{shenouda2008design} and \cite{vucic2009robust}, imperfect CSI can be described by two approaches: probability model and bounded model. When the CSI impairment is dominated by channel estimation errors, the probability model applies. On the other hand, the bounded model is more suitable when quantization errors are the dominant source of CSI uncertainty. In \cite{shenouda2008design} and \cite{zhang2008statistically}, transceiver designs were investigated under the probability model for CSI impairment, while in \cite{vucic2009robust, tajer2011robust, hanif2013efficient, xu2015robust}, the bounded CSI impairment model was adopted. Specifically, \cite{vucic2009robust} studied the mean-squared error (MSE) optimization problem for a single-cell MIMO system, \cite{tajer2011robust, hanif2013efficient} aimed to maximize the worst-case sum SE of a multi-cell network via robust beamforming, and an extended worst-case sum energy efficiency (EE) maximization problem was investigated in \cite{xu2015robust}.

All references \cite{shenouda2008design, zhang2008statistically, vucic2009robust, tajer2011robust, hanif2013efficient, xu2015robust} considered robust transmission design (RTD) problems for the conventional cellular networks. According to our survey, only a few works have studied the RTD problem for single-cell D2D underlaid cellular systems \cite{tang2013outage, memmi2017power, fu2013robust, rahman2015robust}. In \cite{tang2013outage} and \cite{memmi2017power}, the probability model was adopted to characterize the CSI impairment. Specifically, \cite{tang2013outage} aimed to maximize the signal-to-interference-plus-noise ratio (SINR) of the D2D link while guaranteeing the outage probability of the cellular link not exceeding a threshold. However, the considered network was simply composed of a cellular link and a D2D link. In \cite{memmi2017power}, a network with one single-antenna CU and several single-antenna D2D pairs was considered. Centralized and distributed algorithms were proposed to maximize the coverage probability. In \cite{fu2013robust} and \cite{rahman2015robust}, the bounded CSI impairment model was adopted. In particular, \cite{fu2013robust} considered a similar simple network as \cite{tang2013outage} and designed a null-space based robust interference avoiding strategy. Reference \cite{rahman2015robust} considered a D2D underlaid system with one multi-antenna CU and several multi-antenna D2D pairs, and aimed for SINR fairness among D2D users by designing robust transceivers.

To the best of the authors' knowledge, the RTD for a multi-cell D2D underlaid system with multiple CUs and multiple D2D pairs in each cell has not been studied. Hence, it is considered in this paper. In order to increase system SE, it is assumed that all BSs are equipped with multiple antennas \cite{pan2016pricing, pan2017joint}. Since uplink spectrum is often underutilized comparing to that of downlink spectrum in cellular systems \cite{wellens2007evaluation, zulhasnine2010efficient}, uplink resource sharing is thus assumed. According to \cite{jungnickel2014role, rao2013csi}, in the next-generation cellular communication system with advanced estimation schemes, quantization errors will become the main source of CSI uncertainties. Therefore, the bounded model is adopted to characterize the CSI impairment. The main contributions of this paper are summarized as follows:

\hangafter=1
\setlength{\hangindent}{1.9em}
$\bullet$ Different from \cite{tang2013outage, fu2013robust} and \cite{rahman2015robust}, which focused on optimizing the performance of D2D links, considering the higher priorities of CUs, this paper aims to maximize the worst-case sum rate of both cellular and D2D links under CSI impairment, and the interference from all D2D transmitters (D2D-Txs) to each BS is assumed to be power-limited to protect cellular communication. Since a multi-cell network is considered with each cell consisting of a multi-antenna BS, multiple CUs and D2D pairs, joint optimization of the transmit powers of all transmitters and receive filters of all BSs is required, which is more complex. Hence, the algorithms developed in the existing literature cannot be applied directly.

\hangafter=1
\setlength{\hangindent}{1.9em}
$\bullet$ To solve the formulated non-convex problem, the objective function is first analyzed. It is shown that the worst-case SINR of each D2D link can be obtained by independently finding the worst-case (smallest) numerator term and the worst-case (largest) denominator terms, while that of cellular links cannot be analogously found since channel estimation error vectors of cellular links are coupled in SINR expressions. Hence, a precise expression of the objective function cannot be obtained, and it is difficult to solve the original problem.

\hangafter=1
\setlength{\hangindent}{1.9em}
$\bullet$ In order to make the complicated problem tractable, a lower bound on the original objective function is derived by applying the relationship between the minimum mean square error (MMSE) and the SINR of received signals. This lower bound is then maximized subject to the same constraints instead of solving the primal problem. By using the definition of MSE and reformulating the objective function, the resulted problem can be decomposed into multiple semidefinite programming (SDP) subproblems, which are convex and have computationally efficient solutions. An iterative RTD algorithm is then proposed to obtain a suboptimal solution.

\hangafter=1
\setlength{\hangindent}{1.9em}
$\bullet$ In the simulation part, the performance of the proposed RTD algorithm is illustrated and compared in terms of sum SE. It is shown that D2D communication can significantly increase the performance of the conventional cellular system while causing tolerable interference to CUs. In addition, compared with the non-robust transmission design scheme, which takes the estimated channel as the true channel, the sum SE of the system can be greatly increased by the proposed RTD algorithm.

Note that for the sake of simplicity, it is assumed that the RB allocation, i.e., the matching between CUs and DUs, has been predetermined. This paper mainly focuses on the RTD for mobile equipments using the same RB. Obviously, the results obtained in this paper can be applied straightforwardly to a D2D underlaid massive MIMO system where all users use the same RB for transmission \cite{lin2015interplay, xu2017pilot}.

The rest of this paper is organized as follows. In Section \ref{System_Model_and_Problem_Formulation}, a multi-cell D2D underlaid cellular system and problem formulation are presented. In Section \ref{Problem_Analysis_and_Robust_Optimization}, the considered sum SE maximization problem under CSI impairment is first analyzed and an RTD algorithm is then proposed to solve the problem. Finally, numerical verifications are presented in Section \ref{Simulation_Results} before concluding remarks in Section \ref{Conclusions}.

This paper follows commonly used notations. $\mathbb R$ and $\mathbb C$ denote the real space and the complex space, respectively. The boldface upper (lower) case letters are used to denote matrices (vectors). ${\bm I}_M$ stands for the $M \times M$ dimensional identity matrix and $\bm 0$ denotes the all-zero vector or matrix. `` $\setminus$ " represents the set subtraction operation. Superscript $(\cdot)^H$ denotes the conjugated-transpose operation and ${\mathbb E}\{\cdot\}$ denotes the expectation operation. $\left\| {\bm a} \right\|$ is used to denote the Euclidean norm of vector $\bm a$.
\section{System Model and Problem Formulation}
\label{System_Model_and_Problem_Formulation}
\subsection{System Model}

\begin{figure}
  \centering
  \includegraphics[scale=0.31]{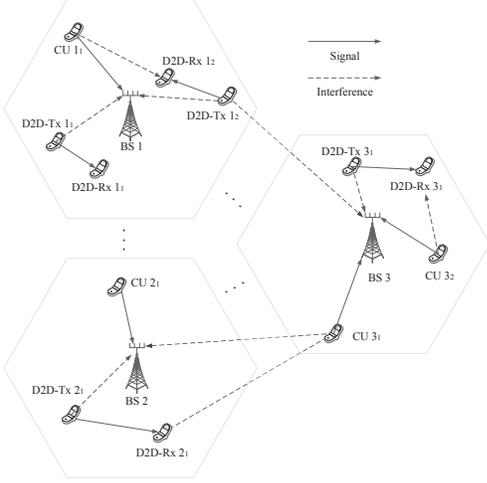}
  \vspace{-0.8em}
  \caption{An example of the considered system diagram.}
  \label{Fig.1}
\end{figure}

As illustrated in Fig. \ref{Fig.1}, this paper considers an $L$-cell interference network with a BS, $M$ CUs and $N$ D2D pairs in each cell while sharing the same uplink RB for transmission.\footnote{Note that the number of CUs and D2D pairs in each cell can be different although the same $M$ and $N$ are used for notational brevity.} To increase the system SE, each BS is assumed to have $B$ antennas to exploit higher spatial degrees of freedom. Each mobile user is equipped with one antenna. For the sake of clarity, only part of the interference signals are depicted in Fig. \ref{Fig.1}. In order to simultaneously communicate with the CUs in each cell, it is assumed that the number of antennas at each BS is not less than that of CUs in the corresponding cell, i.e., $B \geq M$. Let ${\cal L}$, ${\cal C}$ and ${\cal D}$ represent the sets of all BSs, CUs and D2D pairs, respectively. Denote the $m$th cellular user in cell $l$ by CU $l_m$ and the $n$th D2D pair in cell $l$ by $l_n$. Then, the $B\times 1$ dimensional received data vector of BS $l$ can be written as
\begin{equation}
{\bm y}_{l}^{({\text c})} = \!\!\sum\limits_{k_t \in {\cal C}} \!\!\sqrt {p_{k_t}^{({\text c})}} {\bm h}_{k_t,l}^{({\text c})} x_{k_t}^{({\text c})} + \!\!\sum\limits_{k_s \in {\cal D}} \!\!\sqrt {p_{k_s}^{({\text d})}} {\bm h}_{k_s,l}^{({\text d})} x_{k_s}^{({\text d})} + {\bm z}_{l}^{({\text c})},
\label{Received_signal_BS}
\end{equation}
where $p_{k_t}^{({\text c})}$ and $x_{k_t}^{({\text c})}$ denote the transmit power and the zero-mean unit-variance data symbol of CU $k_t$, respectively. ${\bm h}_{k_t,l}^{({\text c})} \in {\mathbb C}^{B\times 1}$ is the channel vector from CU $k_t$ to BS $l$, which accounts for large-scale fading (including path loss and shadow fading) and small-scale fading. $p_{k_s}^{({\text d})}$, $x_{k_s}^{({\text d})}$ and ${\bm h}_{k_s,l}^{({\text d})}$ are similarly defined for D2D-Tx $k_s$. $\bm z_{l}^{({\text c})} \in {\mathbb C}^{B\times 1}$ is the zero-mean circularly symmetric complex white Gaussian noise with covariance $N_0\bm I_B$, i.e., $\bm z_{l}^{({\text c})} \sim {\cal CN}(0,N_0 \bm I_B)$.
Similarly, the received signal of D2D receiver (D2D-Rx) $l_n$ is given by
\begin{equation}
y_{l_n}^{({\text d})} = \!\!\sum\limits_{k_t \in {\cal C}} \!\!\sqrt {p_{k_t}^{({\text c})}} g_{k_t,l_n}^{({\text c})} x_{k_t}^{({\text c})} + \!\!\sum\limits_{k_s \in {\cal D}}\!\! \sqrt {p_{k_s}^{({\text d})}} g_{k_s,l_n}^{({\text d})} x_{k_s}^{({\text d})} + z_{l_n}^{({\text d})},
\label{Received_signal_DU}
\end{equation}
where $g_{k_t,l_n}^{({\text c})}$, $g_{k_s,l_n}^{({\text d})}$ respectively denote the channel coefficient from CU $k_t$ and D2D-Tx $k_s$ to D2D-Rx $l_n$, and $z_{l_n}^{({\text d})}$ is the zero-mean circularly symmetric complex white Gaussian noise with variance $N_0$.

Let ${\bm w}_{l_m} \in {\mathbb C}^{B\times 1}$ denote the unit-norm receive beamforming vector adopted by BS $l$ for detecting signal $x_{l_m}^{({\text c})}$. Then, according to (\ref{Received_signal_BS}) and \cite{xu2017pilotTVT}, the post-processing SINR of CU $l_m$ can be written as
\begin{equation}
{\text {SINR}}_{l_m}^{({\text c})} = \frac{p_{l_m}^{({\text c})} \left|{\bm w}_{l_m}^H {\bm h}_{l_m,l}^{({\text c})} \right|^2}{{\bm w}_{l_m}^H {\bm G}_{l_m} {\bm w}_{l_m}},
\label{SINR_CU}
\end{equation}
where ${\bm G}_{l_m}$ denotes the interference plus noise covariance matrix and is give by
{\setlength\arraycolsep{2pt}
\begin{eqnarray}
\!\!\!\!\!&&{\bm G}_{l_m}\!\!=\!{\mathbb E}\!\left\{\!\!\left(\!{\bm y}_{l}^{({\text c})}\!- \!\sqrt {p_{l_m}^{({\text c})}} {\bm h}_{l_m,l}^{({\text c})} x_{l_m}^{({\text c})}\!\right)^{\!H} \!\!\left(\!{\bm y}_{l}^{({\text c})}\!-\! \sqrt {p_{l_m}^{({\text c})}} {\bm h}_{l_m,l}^{({\text c})} x_{l_m}^{({\text c})}\!\right) \!\!\right\}\nonumber\\
\!\!\!\!\!&&=\!\! \sum\limits_{k_t \in {\cal C} \setminus l_m} \!\!\!p_{k_t}^{({\text c})} {\bm h}_{k_t,l}^{({\text c})} \!\left(\!{\bm h}_{k_t,l}^{({\text c})}\!\right)^{\!H} \!\!+\!\! \sum\limits_{k_s\in {\cal D}} \!p_{k_s}^{({\text d})} \! {\bm h}_{k_s,l}^{({\text d})} \!\left(\!{\bm h}_{k_s,l}^{({\text d})}\!\right)^{\!H} \!\!+ N_0 {\bm I}_B.
\label{covariance_matrix}
\end{eqnarray}}
\!\!\!The second equality of (\ref{covariance_matrix}) holds because it is assumed that each transmitter independently sends zero-mean unit-variance data symbols. Analogously, from (\ref{Received_signal_DU}), the received SINR of D2D-Rx $l_n$ can be expressed as
\begin{equation}
{\text {SINR}}_{l_n}^{({\text d})} \!=\! \frac{p_{l_n}^{({\text d})} \left|g_{l_n,l_n}^{({\text d})}\right|^2}{\sum\limits_{k_t \in {\cal C}} p_{k_t}^{({\text c})} \!\left|g_{k_t,l_n}^{({\text c})}\right|^2 \!+\! \sum\limits_{k_s \in {\cal D} \setminus l_n}\! p_{k_s}^{({\text d})} \!\left|g_{k_s,l_n}^{({\text d})}\right|^2 \!+\! N_0}.
\label{SINR_DU}
\end{equation}

As mentioned in Section \ref{Introduction}, it is difficult for the BSs to obtain the perfect CSI of all channels, and with advanced estimation schemes, quantization errors will become the main source of CSI uncertainties in the next-generation cellular communication system \cite{jungnickel2014role, rao2013csi}. Therefore, it is assumed that the BSs only have partial CSI, and the bounded model is adopted to characterize the CSI impairment. Denote the imperfect CSI of ${\bm h}_{k_t,l}^{({\text c})}$, ${\bm h}_{k_s,l}^{({\text d})}$, $g_{k_t,l_n}^{({\text c})}$ and $g_{k_s,l_n}^{({\text d})}$ by ${\tilde {\bm h}}_{k_t,l}^{({\text c})}$, ${\tilde {\bm h}}_{k_s,l}^{({\text d})}$, ${\tilde g}_{k_t,l_n}^{({\text c})}$ and ${\tilde g}_{k_s,l_n}^{({\text d})}$, and the corresponding CSI errors by ${\bm \Delta}_{k_t,l}^{({\text c})}$, ${\bm \Delta}_{k_s,l}^{({\text d})}$, $\delta_{k_t,l_n}^{({\text c})}$ and $\delta_{k_s,l_n}^{({\text d})}$, i.e.,
{\setlength\arraycolsep{2pt}
\begin{eqnarray}
&&{\bm \Delta}_{k_t,l}^{({\text c})}={\bm h}_{k_t,l}^{({\text c})}-{\tilde {\bm h}}_{k_t,l}^{({\text c})}, ~\forall ~k_t \in {\cal C},~l \in {\cal L},\nonumber\\
&&{\bm \Delta}_{k_s,l}^{({\text d})}={\bm h}_{k_s,l}^{({\text d})}-{\tilde {\bm h}}_{k_s,l}^{({\text d})}, ~\forall ~k_s \in {\cal D},~l \in {\cal L} ,\nonumber\\
&&\delta_{k_t,l_n}^{({\text c})}=g_{k_t,l_n}^{({\text c})}-{\tilde g}_{k_t,l_n}^{({\text c})}, ~\forall ~k_t \in {\cal C},~l_n \in {\cal D}, \nonumber\\
&&\delta_{k_s,l_n}^{({\text d})}=g_{k_s,l_n}^{({\text d})}-{\tilde g}_{k_s,l_n}^{({\text d})}, ~\forall ~k_s \in {\cal D},~l_n \in {\cal D}.
\label{CSI_error}
\end{eqnarray}}
\!\!\!For the bounded model, the errors in (\ref{CSI_error}) satisfy
{\setlength\arraycolsep{2pt}
\begin{subequations}
\begin{align}
&\left\| {\bm \Delta}_{k_t,l}^{({\text c})} \right\| \leq \varepsilon_{k_t,l}^{({\text c})}, ~\forall~ k_t\in {\cal C},~l \in {\cal L},\label{estimation_error_a}\\
&\left\| {\bm \Delta}_{k_s,l}^{({\text d})} \right\| \leq \varepsilon_{k_s,l}^{({\text d})}, ~\forall~ k_s \in {\cal D},~l \in {\cal L},\label{estimation_error_b}\\
&\left| \delta_{k_t,l_n}^{({\text c})} \right| \leq \epsilon_{k_t,l_n}^{({\text c})}, ~\forall~ k_t\in {\cal C},~l_n \in {\cal D},\label{estimation_error_c}\\
&\left| \delta_{k_s,l_n}^{({\text d})} \right| \leq \epsilon_{k_s,l_n}^{({\text d})}, ~\forall~ k_s \in {\cal D},~l_n \in {\cal D},\label{estimation_error_d}
\end{align}
\label{estimation_error}
\end{subequations}}
\!\!\!where $\varepsilon_{k_t,l}^{({\text c})}$, $\varepsilon_{k_s,l}^{({\text d})}$, $\epsilon_{k_t,l_n}^{({\text c})}$ and $\epsilon_{k_s,l_n}^{({\text d})}$ represent the corresponding CSI error bounds.
\subsection{Problem Formulation}
This paper aims to maximize the worst-case sum SE of the cooperative multi-cell network under imperfect CSI. From (\ref{SINR_CU}) and (\ref{SINR_DU}), the network throughput can be expressed as
{\setlength\arraycolsep{2pt}
\begin{eqnarray}
R &=& \sum\limits_{l_m \in {\cal C}} R_{l_m}^{({\text c})} + \sum\limits_{l_n \in {\cal D}} R_{l_n}^{({\text d})} \nonumber\\
&=& \sum\limits_{l_m \in {\cal C}} \log_2\left( 1 \!+\! {\text {SINR}}_{l_m}^{({\text c})} \right) \!+\! \sum\limits_{l_n \in {\cal D}} \log_2\left( 1 \!+\! {\text {SINR}}_{l_n}^{({\text d})} \right),\quad
\label{sum_rate}
\end{eqnarray}}
\!\!\!where $R_{l_m}^{({\text c})}$ and $R_{l_n}^{({\text d})}$ respectively denote the throughput of cellular link $l_m$ and D2D link $l_n$. In order to guarantee the QoS of cellular links, assume that the interference signal from all D2D-Txs to each BS is power-limited \cite{kim2013decentralized}, i.e.,
\begin{equation}
\sum\limits_{k_s\in {\cal D}} p_{k_s}^{({\text d})} \left\|{\bm h}_{k_s,l}^{({\text d})} \right\|^2 \leq a_l, ~\forall~ l \in {\cal L},
\label{interference_threshold}
\end{equation}
where $a_l$ is the maximum interference threshold to protect CUs in the $l$th cell. Since the imperfect CSI is considered, an accurate value of the left-hand side term of (\ref{interference_threshold}) becomes intractable. Therefore, (\ref{interference_threshold}) is rewritten in a stricter form in the sequel. Beforehand, Lemma \ref{lemma_0} is first given which is a simple extension of a result in \cite{vorobyov2004adaptive} and can be readily proven.
\begin{lemma}
For any given ${\tilde {\bm h}} \in {\mathbb C}^{B\times 1}$, $\varepsilon \in R_+$ and the uncertainty region $\Omega \!=\! \left\{{\bm \Delta} | {\bm \Delta} \in {\mathbb C}^{B\times 1}, \left\| {\bm \Delta} \right\| \leq \varepsilon \right\}$, the following results hold
{\setlength\arraycolsep{2pt}
\begin{eqnarray}
&& \mathop {\min }\limits_{{\bm \Delta} \in \Omega} \left\|{\tilde {\bm h}} + {\bm \Delta}\right\|^2 =  \left[ \left( \left\|\tilde {\bm h}\right\| - \varepsilon \right)^+\right]^2, \nonumber\\
&& \mathop {\max }\limits_{{\bm \Delta} \in \Omega} \left\|{\tilde {\bm h}} + {\bm \Delta}\right\|^2 = \left( \left\|\tilde {\bm h}\right\| + \varepsilon \right)^2,
\end{eqnarray}}
\!\!whose optimal solutions are respectively given by
{\setlength\arraycolsep{2pt}
\begin{eqnarray}
&& \arg \mathop {\min }\limits_{{\bm \Delta} \in \Omega} \left\|{\tilde {\bm h}} + {\bm \Delta}\right\|^2 = -{\tilde {\bm h}} \times \min \left\{1, \frac{\varepsilon }{\left\|\tilde {\bm h}\right\|}\right\}, \nonumber\\
&& \arg \mathop {\max }\limits_{{\bm \Delta} \in \Omega} \left\|{\tilde {\bm h}} + {\bm \Delta}\right\|^2 = \frac{\varepsilon {\tilde {\bm h}}}{\left\|\tilde {\bm h}\right\|}.
\end{eqnarray}}
\label{lemma_0}
\end{lemma}

Now by recalling (\ref{CSI_error}) and (\ref{estimation_error}), and invoking the result of the lemma above, a stricter form of (\ref{interference_threshold}) is given by
\begin{equation}
\sum\limits_{k_s\in {\cal D}} p_{k_s}^{({\text d})} \rho_{k_s,l} \leq a_l, ~\forall~ l \in {\cal L},
\label{SINR_reduce_strict}
\end{equation}
where $\rho_{k_s,l} = \left( \left\| {\tilde {\bm h}}_{k_s,l}^{({\text d})} \right\| + \varepsilon_{k_s,l}^{({\text d})}\right)^2$.

Let ${\bm \Delta}$ and ${\bm \delta}$ respectively denote the sets of CSI errors from all transmitters to all BSs and D2D-Rxs, i.e.,
{\setlength\arraycolsep{2pt}
\begin{eqnarray}
&&{\bm \Delta} = \left[{\bm \Delta}_1, \cdots, {\bm \Delta}_L \right], \nonumber\\
&&{\bm \delta} = \left[{\bm \delta}_{1_1}, \cdots, {\bm \delta}_{1_N}, \cdots, {\bm \delta}_{L_N} \right],
\label{set0}
\end{eqnarray}}
\!\!\!where
{\setlength\arraycolsep{2pt}
\begin{eqnarray}
&&{\bm \Delta}_l = \left[ {\bm \Delta}_{1_1,l}^{({\text c})}, \cdots, {\bm \Delta}_{1_M,l}^{({\text c})}, \cdots, {\bm \Delta}_{L_M,l}^{({\text c})}, {\bm \Delta}_{1_1,l}^{({\text d})}, \cdots,\right. \nonumber\\
&&\quad\quad\quad \left.{\bm \Delta}_{1_N,l}^{({\text d})}, \cdots,{\bm \Delta}_{L_N,l}^{({\text d})}\right], ~\forall~ l \in {\cal L}, \nonumber\\
&&{\bm \delta}_{l_n} = \left(\delta_{1_1,l_n}^{({\text c})}, \cdots, \delta_{1_M,l_n}^{({\text c})}, \cdots, \delta_{L_M,l_n}^{({\text c})}, \delta_{1_1,l_n}^{({\text d})}, \cdots,\right.  \nonumber\\
&&\quad\quad\quad \left.\delta_{1_N,l_n}^{({\text d})}, \cdots, \delta_{L_N,l_n}^{({\text d})}\right)^T, ~\forall~ l_n \in {\cal D}.
\label{set1}
\end{eqnarray}}
\!\!Since CSI errors are unknown and norm-bounded by (\ref{estimation_error}), there always exists a worst case (denote the corresponding CSI errors as ${\bm \Delta}^*$ and ${\bm \delta}^*$) such that ${\bm \Delta}^*$ and ${\bm \delta}^*$ satisfy (\ref{estimation_error}), and the system outputs the worst-case SE $R^*$. Then, the worst-case SE maximization problem can be formulated as
{\setlength\arraycolsep{2pt}
\begin{subequations}
\begin{align}
\mathop {\max }\limits_{ {\bm W}, {\bm p}}  \quad& R^*  \label{optimize_problem_a}\\
\text{s.t.} \quad\; & 0 \leq p_{l_m}^{({\text c})} \leq P_{l_m}^{({\text c})}, ~\forall~ l_m \in {\cal C},\label{optimize_problem_b}\\
& 0 \leq p_{l_n}^{({\text d})} \leq P_{l_n}^{({\text d})}, ~\forall~ l_n \in {\cal D}, \label{optimize_problem_c}\\
& \sum\limits_{k_s\in {\cal D}} p_{k_s}^{({\text d})} \rho_{k_s,l} \leq a_l, ~\forall~ l \in {\cal L}, \label{optimize_problem_d}
\end{align}
\label{optimize_problem}
\end{subequations}}
\!\!\!\!\!where $P_{l_m}^{({\text c})}$ and $P_{l_n}^{({\text d})}$ denote the maximum transmit power of CU $l_m$ and D2D-Tx $l_n$, respectively. ${\bm W}$ and ${\bm p}$ are defined as follows
{\setlength\arraycolsep{2pt}
\begin{eqnarray}
&&{\bm W}=\left[{\bm W}_1,\cdots,{\bm W}_L\right], \nonumber\\
&&{\bm p} = \left[ {\bm p}^{({\text c})}; {\bm p}^{({\text d})}\right],
\label{set}
\end{eqnarray}}
\!\!\!where
{\setlength\arraycolsep{2pt}
\begin{eqnarray}
&&{\bm W}_l = \left[{\bm w}_{l_1}, \cdots, {\bm w}_{l_M}\right], ~\forall~ l \in {\cal L},\nonumber\\
&&{\bm p}^{({\text c})} = \left(p_{1_1}^{({\text c})}, \cdots, p_{1_M}^{({\text c})}, \cdots, p_{L_M}^{({\text c})}\right)^T, \nonumber\\
&&{\bm p}^{({\text d})} = \left(p_{1_1}^{({\text d})}, \cdots, p_{1_N}^{({\text d})}, \cdots, p_{L_N}^{({\text d})}\right)^T.
\label{set2}
\end{eqnarray}}

Note that the actual CSI errors are not variables and can not be optimized. However, by using the transmit power
vector and receive beamforming vectors designed for the worst case, a better system throughput can be obtained than the nonrobust design scheme, which simply takes the estimated channel as the true channel. Due to the following two challenges, it is difficult to directly solve problem (\ref{optimize_problem}): First, CSI error vectors are coupled in the SINR calculation, making it hard to obtain an explicit expression of the objective function $R^*$; Second, the fractional form of the SINR expressions and the $\log(\cdot)$ operation make the objective function nonconvex. To handle this problem, an alternative algorithm will be proposed in the following section.
\section{Problem Analysis and Robust Optimization}
\label{Problem_Analysis_and_Robust_Optimization}
In this section, the worst-case SE maximization problem (\ref{optimize_problem}) is investigated. As stated above, it is difficult to directly solve (\ref{optimize_problem}). Therefore, in the following of this section, problem (\ref{optimize_problem}) is first analyzed and transformed to a more tractable form. Then, an alternative algorithm is proposed to solve it.
\subsection{Problem Analysis}
\label{Problem_Analysis}
From (\ref{SINR_DU}), it can be found that each channel coefficient from a mobile transmitter to a D2D-Rx appears only once in either the numerator or denominator of ${\text {SINR}}_{l_n}^{({\text d})}, ~\forall~ l_n \in {\cal D}$, and all channel coefficients $g_{k_t,l_n}^{({\text c})}, g_{k_s,l_n}^{({\text d})}, ~\forall~ k_t \in {\cal C},~ k_s,~ l_n \in {\cal D}$ are independent of each other. As a result, the worst-case ${\text {SINR}}_{l_n}^{({\text d})}$ can be obtained by decoupling it into finding the worst-case (smallest) numerator term and the worst-case (largest) denominator terms, which can be implemented based on Lemma \ref{lemma_0}. Define the worst-case ${\text {SINR}}_{l_n}^{({\text d})}$ by ${\text {sinr}}_{l_n}^{({\text d})}\triangleq \mathop {\min }\limits_{{\bm \delta}_{l_n}} {\text {SINR}}_{l_n}^{({\text d})}$. It follows that
{\setlength\arraycolsep{2pt}
\begin{eqnarray}
\!\!\!\!\!\!&&{\text {sinr}}_{l_n}^{({\text d})} = \mathop {\min }\limits_{{\bm \delta}_{l_n}} \frac{p_{l_n}^{({\text d})} \left|g_{l_n,l_n}^{({\text d})}\right|^2}{\sum\limits_{k_t \in {\cal C}} \!p_{k_t}^{({\text c})} \left|g_{k_t,l_n}^{({\text c})}\right|^2 \!+\! \sum\limits_{k_s \in {\cal D} \setminus l_n} \!p_{k_s}^{({\text d})} \left|g_{k_s,l_n}^{({\text d})}\right|^2 \!+\! N_0}\nonumber\\
\!\!\!\!\!\!&&= \frac{p_{l_n}^{({\text d})} \mathop {\min }\limits_{\delta_{l_n,l_n}^{({\text d})}} \left|g_{l_n,l_n}^{({\text d})}\right|^2}{\sum\limits_{k_t \in {\cal C}} \!p_{k_t}^{({\text c})} \mathop {\max }\limits_{\delta_{k_t,l_n}^{({\text c})}} \left|g_{k_t,l_n}^{({\text c})}\right|^2 \!+\! \sum\limits_{k_s \in {\cal D} \setminus l_n} \!p_{k_s}^{({\text d})} \mathop {\max }\limits_{\delta_{k_s,l_n}^{({\text d})}}\left|g_{k_s,l_n}^{({\text d})}\right|^2 \!+\! N_0}\nonumber\\
\!\!\!\!\!\!&&= \frac{p_{l_n}^{({\text d})} \left|{\bar g}_{l_n,l_n}^{({\text d})}\right|^2}{\sum\limits_{k_t \in {\cal C}} \!p_{k_t}^{({\text c})}\! \left|{\bar g}_{k_t,l_n}^{({\text c})}\right|^2 \!\!+\!\! \sum\limits_{k_s \in {\cal D} \setminus l_n} \!\!p_{k_s}^{({\text d})}\! \left|{\bar g}_{k_s,l_n}^{({\text d})}\right|^2 \!\!+\! N_0},
\label{worst_SINR_DU}
\end{eqnarray}}
\!\!where Lemma \ref{lemma_0} is used in the last equality, and ${\bar g}_{l_n,l_n}^{({\text d})}$, ${\bar g}_{k_t,l_n}^{({\text c})}$ and ${\bar g}_{k_s,l_n}^{({\text d})}$ are given by
{\setlength\arraycolsep{2pt}
\begin{eqnarray}
{\bar g}_{l_n,l_n}^{({\text d})} &=& {\tilde g}_{l_n,l_n}^{({\text d})} - {\tilde g}_{l_n,l_n}^{({\text d})} \times \min \left\{1, \frac{\epsilon_{l_n,l_n}^{({\text d})}}{\left|{\tilde g}_{l_n,l_n}^{({\text d})}\right|}\right\}, ~\forall~ l_n \in {\cal D},\nonumber\\
{\bar g}_{k_t,l_n}^{({\text c})} &=& {\tilde g}_{k_t,l_n}^{({\text c})} + \frac{\epsilon_{k_t,l_n}^{({\text c})} {\tilde g}_{k_t,l_n}^{({\text c})}}{\left|{\tilde g}_{k_t,l_n}^{({\text c})}\right|}, ~\forall~ k_t \in {\cal C},~ l_n \in {\cal D},\nonumber\\
{\bar g}_{k_s,l_n}^{({\text d})} &=& {\tilde g}_{k_s,l_n}^{({\text d})} + \frac{\epsilon_{k_s,l_n}^{({\text d})} {\tilde g}_{k_s,l_n}^{({\text d})}}{\left|{\tilde g}_{k_s,l_n}^{({\text d})}\right|}, ~\forall~ k_s,~ l_n \in {\cal D},~ k_s\neq l_n.
\label{worst_channel_DU}
\end{eqnarray}}

In contrast, for cellular links, since the CUs in each cell simultaneously communicate with the corresponding BS on the same RB, it can be found from (\ref{SINR_CU}) and (\ref{covariance_matrix}) that the channel vector ${\bm h}_{l_m,l}^{({\text c})}$ appears not only in the numerator term of $\text {SINR}_{l_m}^{({\text c})}$ but also in the denominator terms of $\text {SINR}_{k_t}^{({\text c})}, ~\forall~ k_t \in {\cal C}_l \setminus l_m$, where ${\cal C}_l$ denotes the set of all CUs in cell $l$. Therefore, the worst-case $\text {SINR}_{l_m}^{({\text c})}$ cannot be obtained by independently finding the worst-case numerator and the worst-case terms in the denominator. For example, in cell $3$ of Fig. \ref{Fig.1}, CU $3_1$ and CU $3_2$ simultaneously communicate with BS $3$. Hence, ${\bm h}_{3_1,3}^{({\text c})}$ (${\bm h}_{3_2,3}^{({\text c})}$) appears both in the numerator term of $\text {SINR}_{3_1}^{({\text c})}$ ($\text {SINR}_{3_2}^{({\text c})}$) and in the denominator terms of $\text {SINR}_{3_2}^{({\text c})}$ ($\text {SINR}_{3_1}^{({\text c})}$) due to co-channel interference. It is thus difficult to directly obtain the worst-case $\text {SINR}_{3_1}^{({\text c})}$ and $\text {SINR}_{3_2}^{({\text c})}$.

To deal with the SINR expressions in fractional form, the relationship between the MMSE and the pre-processing SINR was applied in \cite{tajer2011robust} and \cite{xu2015robust}, which considered a downlink multi-cell system. In the following, it is shown that a similar relationship also holds for a multi-cell uplink system.

\begin{theorem}
In a D2D underlaid multi-cell uplink system, if the MMSE filter ${\bm w}_{l_m}^{\text {MMSE}}$ is adopted by BS $l$ for detecting $x_{l_m}^{({\text c})}$, denote the MMSE of cellular link $l_m$ by ${\text {MMSE}}_{l_m}^{({\text c})}$. Then, the following relationship exists
\begin{equation}
{\text {MMSE}}_{l_m}^{({\text c})}=\frac{1}{1+\text {SINR}_{l_m}^{({\text c})}}.
\label{MMSE_SINR_CU}
\end{equation}
As for D2D link $l_n$, denote the worst-case MMSE by ${\text {MMSE}}_{l_n}^{({\text d})}$. Then, an analogous relationship between ${\text {MMSE}}_{l_n}^{({\text d})}$ and $\text {sinr}_{l_n}^{({\text d})}$ can also be obtained, i.e.,
\begin{equation}
{\text {MMSE}}_{l_n}^{({\text d})}=\frac{1}{1+\text {sinr}_{l_n}^{({\text d})}}.
\label{MMSE_SINR_DU}
\end{equation}
\label{theorem_0}
\end{theorem}
\itshape \textbf{Proof:}  \upshape
See Appendix A.
\hfill $\Box$

Based on (\ref{worst_SINR_DU}) and applying Theorem \ref{theorem_0}, the objective function of (\ref{optimize_problem}), i.e., $R^*$ can be equivalently rewritten as\footnote{ Note that for the convenience of the following analysis, $\log(\cdot)$ is replaced with $\ln(\cdot)$ in (\ref{equivalent_R}).}
{\setlength\arraycolsep{2pt}
\begin{eqnarray}
\!\!\!\!\!\!&& \mathop {\min }\limits_{\bm \Delta} \sum\limits_{l_m \in {\cal C}} \ln\left( 1 + {\text {SINR}}_{l_m}^{({\text c})} \right) + \sum\limits_{l_n \in {\cal D}} \ln\left( 1 + {\text {sinr}}_{l_n}^{({\text d})} \right)  \nonumber\\
\!\!\!\!\!\!&&= -\sum\limits_{l=1}^L \mathop {\max }\limits_{{\bm \Delta}_l} \sum\limits_{m=1}^M \ln {\text {MMSE}}_{l_m}^{({\text c})} - \sum\limits_{l_n \in {\cal D}}  \ln {\text {MMSE}}_{l_n}^{({\text d})} \nonumber\\
\!\!\!\!\!\!&&= -\sum\limits_{l=1}^L \mathop {\max }\limits_{{\bm \Delta}_l} \mathop {\min }\limits_{{\bm W}_l} \sum\limits_{m=1}^M  \ln {\text {MSE}}_{l_m}^{({\text c})} - \sum\limits_{l_n \in {\cal D}} \mathop {\min }\limits_{f_{l_n}} \ln {\text {MSE}}_{l_n}^{({\text d})},\quad
\label{equivalent_R}
\end{eqnarray}}
\!\!\!where $f_{l_n} \!\in\! {\mathbb C}\backslash \{0\}$ is the single-tap receive equalizer at D2D-Rx $l_n$. ${\text {MSE}}_{l_m}^{({\text c})}$ and ${\text {MSE}}_{l_n}^{({\text d})}$ respectively denote the MSEs of cellular link $l_m$ and D2D link $l_n$. The first equality of (\ref{equivalent_R}) holds because ${\bm \Delta}_l, ~\forall~ l \in {\cal L}$ have independent uncertainties. According to (\ref{equivalent_R}), the worst-case SE maximization problem (\ref{optimize_problem}) can be equivalently rewritten as
{\setlength\arraycolsep{2pt}
\begin{eqnarray}
\!\!\!\mathop {\min }\limits_{\bm p} && \left(\sum\limits_{l=1}^L  \mathop {\max }\limits_{{\bm \Delta}_l} \mathop {\min }\limits_{{\bm W}_l} \sum\limits_{m=1}^M  \ln {\text {MSE}}_{l_m}^{({\text c})} \!+\! \sum\limits_{l_n \in {\cal D}} \!\mathop {\min }\limits_{f_{l_n}} \ln {\text {MSE}}_{l_n}^{({\text d})} \!\right) \label{optimize_problem26_a}\nonumber\\
\text{s.t.} \; && {\text {(\ref{optimize_problem_b})}}\sim {\text {(\ref{optimize_problem_d})}},~{\text {(\ref{estimation_error_a})}},~ {\text {(\ref{estimation_error_b})}}.\label{optimize_problem26_b}
\label{optimize_problem26}
\end{eqnarray}}

From (\ref{Received_signal_BS}) and the definition of MSE, it is known that ${\bm h}_{l_m,l}^{({\text c})}$ appears in all ${\text {MSE}}_{l_m}^{({\text c})}, ~\forall~ l_m \in {\cal C}_l$, making it difficult to obtain a tractable expression of $\sum\limits_{m=1}^M  \ln {\text {MSE}}_{l_m}^{({\text c})}$ due to the $\ln(\cdot)$ operation. In addition, the composite min-max-min optimization makes it more difficult to solve (\ref{optimize_problem26}).

To simplify problem (\ref{optimize_problem26}), recall that for any function $f(x,y)$, the inequality $\mathop {\min }\limits_x \mathop {\max }\limits_y f(x,y) \geq  \mathop {\max }\limits_y \mathop {\min }\limits_x f(x,y)$ always holds \cite{tajer2011robust}. By exchanging the positions of $\max$ and $\min$, an upper bound on the objective function of (\ref{optimize_problem26}) can be obtained and problem (\ref{optimize_problem26}) can be simplified as follows
{\setlength\arraycolsep{2pt}
\begin{eqnarray}
\!\!\!\mathop {\min }\limits_{\bm p} && \left(\sum\limits_{l=1}^L \mathop {\min }\limits_{{\bm W}_l} \mathop {\max }\limits_{{\bm \Delta}_l} \sum\limits_{m=1}^M  \ln {\text {MSE}}_{l_m}^{({\text c})} \!+\! \sum\limits_{l_n \in {\cal D}} \!\mathop {\min }\limits_{f_{l_n}} \ln {\text {MSE}}_{l_n}^{({\text d})} \!\right) \label{optimize_problem2_a}\nonumber\\
\text{s.t.} \; && {\text {(\ref{optimize_problem_b})}}\sim {\text {(\ref{optimize_problem_d})}},~{\text {(\ref{estimation_error_a})}},~ {\text {(\ref{estimation_error_b})}}.
\label{optimize_problem2}
\end{eqnarray}}
\!\!\!For further simplification, motivated by \cite{christensen2008weighted}, the following auxiliary function is introduced to remove the $\ln(\cdot)$ operation in (\ref{optimize_problem2_a})
\begin{equation}
S_l \left(\!{\bm u}_l^{({\text c})}\!\right) \!=\! \mathop {\max }\limits_{{\bm \Delta}_l} \!\sum\limits_{m=1}^M \!\left\{ \exp \left(u_{l_m}^{({\text c})}\!-\!1\right){\text {MSE}}_{l_m}^{({\text c})} \!-\! u_{l_m}^{({\text c})} \right\}, ~\forall~ l\in {\cal L},
\label{auxiliary_CU}
\end{equation}
where ${\bm u}_l^{({\text c})} = \left(u_{l_1}^{({\text c})},\cdots,u_{l_M}^{({\text c})}\right)^T \in {\mathbb R^{M\times 1}}$ is a newly introduced auxiliary vector variable. By checking the first-order optimality condition of (\ref{auxiliary_CU}),
\begin{equation}
\mathop {\min }\limits_{{\bm u}_l^{({\text c})}} S_l \left({\bm u}_l^{({\text c})}\right) = \mathop {\max }\limits_{{\bm \Delta}_l} \sum\limits_{m=1}^M  \ln {\text {MSE}}_{l_m}^{({\text c})}, ~\forall~ l\in {\cal L}.
\label{auxiliary_CU2}
\end{equation}

Similarly, for D2D links, an auxiliary function is introduced, viz.,
\begin{equation}
T_{l_n} \left(u_{l_n}^{({\text d})}\right) = \exp \left(u_{l_n}^{({\text d})} - 1\right) {\text {MSE}}_{l_n}^{({\text d})} - u_{l_n}^{({\text d})}, ~\forall~ l_n \in {\cal D}.
\label{auxiliary_DU}
\end{equation}
It follows
\begin{equation}
\mathop {\min }\limits_{u_{l_n}^{({\text d})}} T_{l_n} \left(u_{l_n}^{({\text d})}\right) = \ln {\text {MSE}}_{l_n}^{({\text d})}, ~\forall~ l_n \in {\cal D},
\label{auxiliary_DU2}
\end{equation}
and the corresponding optimal $u_{l_n}^{({\text d})*}$
\begin{equation}
u_{l_n}^{({\text d})*}= 1- \ln {\text {MSE}}_{l_n}^{({\text d})}, ~\forall~ l_n \in {\cal D}.
\label{auxiliary_DU3}
\end{equation}

Substituting (\ref{auxiliary_CU2}) and (\ref{auxiliary_DU2}) into (\ref{optimize_problem2_a}), and using the independence of ${\bm W}_l$, $f_{l_n}$, ${\bm u}_l^{({\text c})}$ as well as $u_{l_n}^{({\text d})}$, the objective function of problem (\ref{optimize_problem2}) can be rewritten as
{\setlength\arraycolsep{2pt}
\begin{eqnarray}
&& \sum\limits_{l=1}^L \mathop {\min }\limits_{{\bm W}_l} \mathop {\min }\limits_{{\bm u}_l^{({\text c})}} S_l \left({\bm u}_l^{({\text c})}\right) + \sum\limits_{l_n \in {\cal D}} \mathop {\min }\limits_{f_{l_n}} \mathop {\min }\limits_{u_{l_n}^{({\text d})}} T_{l_n} \left(u_{l_n}^{({\text d})}\right)\nonumber\\
& =& \mathop {\min }\limits_{{\bm W},{\bm f},{\bm u}} \left\{ \sum\limits_{l=1}^L S_l \left({\bm u}_l^{({\text c})}\right) + \sum\limits_{l_n \in {\cal D}} T_{l_n} \left(u_{l_n}^{({\text d})}\right) \right\},
\label{objective_function}
\end{eqnarray}}
\!\!\!where ${\bm f}$ and ${\bm u}$ are defined as follows
{\setlength\arraycolsep{2pt}
\begin{eqnarray}
&&{\bm f} = \left(f_{1_1},\cdots,f_{1_N},\cdots, f_{L_N}\right)^T, \nonumber\\
&&{\bm u} = [{\bm u}^{({\text c})};{\bm u}^{({\text d})}], \nonumber\\
&&{\bm u}^{({\text c})} = [{\bm u}_1^{({\text c})};\cdots;{\bm u}_L^{({\text c})}],\nonumber\\
&&{\bm u}^{({\text d})} = \left(u_{1_1}^{({\text d})},\cdots,u_{1_N}^{({\text d})},\cdots, u_{L_N}^{({\text d})}\right)^T.
\end{eqnarray}}
\!\!\!\!Substituting (\ref{auxiliary_CU}) and (\ref{auxiliary_DU}) in (\ref{objective_function}), problem (\ref{optimize_problem2}) can be reformulated as
{\setlength\arraycolsep{2pt}
\begin{eqnarray}
\!\!\!\mathop {\min }\limits_{{\bm p},{\bm W},{\bm f},{\bm u}} && \left\{\sum\limits_{l=1}^L \mathop {\max }\limits_{{\bm \Delta}_l} \sum\limits_{m=1}^M \left[ \exp \left(u_{l_m}^{({\text c})}-1\right){\text {MSE}}_{l_m}^{({\text c})} - u_{l_m}^{({\text c})} \right]\right.\nonumber\\
&& \left.+ \sum\limits_{l_n \in {\cal D}} \left[ \exp \left(u_{l_n}^{({\text d})} - 1\right) {\text {MSE}}_{l_n}^{({\text d})} - u_{l_n}^{({\text d})} \right] \right\} \nonumber\\
\!\!\!\text{s.t.} \;\;\;\, && {\text {(\ref{optimize_problem_b})}}\sim {\text {(\ref{optimize_problem_d})}},~{\text {(\ref{estimation_error_a})}},~ {\text {(\ref{estimation_error_b})}}.
\label{optimize_problem3}
\end{eqnarray}}

Since an explicit expression of the objective function of (\ref{optimize_problem3}) is unavailable, it is difficult to solve this problem. In the following subsection, problem (\ref{optimize_problem3}) is divided into two consecutive parts and an alternative algorithm is proposed to solve it. In the first part, ${\bm W}$, ${\bm f}$ and ${\bm u}$ are optimized for fixed ${\bm p}$, and vice versa in the second part.
\newcounter{TempEqCnt}
\setcounter{TempEqCnt}{\value{equation}}
\setcounter{equation}{37}
\begin{figure*}[hb]
\hrulefill
{\setlength\arraycolsep{2pt}
\begin{eqnarray}
&& \sum\limits_{l_m \in {\cal C}} \left[\left(\gamma_{l_m}^{({\text c})}\right)^2 {\text {MSE}}_{l_m}^{({\text c})} - 2\ln \gamma_{l_m}^{({\text c})} -1 \right] + \sum\limits_{l_n \in {\cal D}} \left[\left(\gamma_{l_n}^{({\text d})}\right)^2 {\text {MSE}}_{l_n}^{({\text d})} - 2\ln \gamma_{l_n}^{({\text d})} -1 \right] \nonumber\\
&=& \sum\limits_{l_m \in {\cal C}} \left\{ \left\| q_{l_m}^{({\text c})} {\bm J}_l^H {\bm h}_{l_m,l}^{({\text c})} - \gamma_{l_m}^{({\text c})} {\bm e}_m\right\|^2 + \sum\limits_{k\neq l} \left\| q_{l_m}^{({\text c})} {\bm J}_k^H {\bm h}_{l_m,k}^{({\text c})}\right\|^2 + \left(q_{l_m}^{({\text c})}\right)^2 \sum\limits_{k_s \in {\cal D}} \left|\gamma_{k_s}^{({\text d})} f_{k_s}^H {\bar g}_{l_m,k_s}^{({\text c})} \right|^2 \right\} \nonumber\\
&+& \sum\limits_{l_n \in {\cal D}} \left\{ \left(q_{l_n}^{({\text d})}\right)^2 \left[\sum\limits_{k=1}^L \left\| {\bm J}_k^H {\bm h}_{l_n,k}^{({\text d})} \right\|^2 + \sum\limits_{k_s \in {\cal D}} \left|\gamma_{k_s}^{({\text d})} f_{k_s}^H {\bar g}_{l_n,k_s}^{({\text d})} \right|^2\right] - 2 q_{l_n}^{({\text d})} \left(\gamma_{l_n}^{({\text d})}\right)^2  {\text {Re}}\left( f_{l_n}^H {\bar g}_{l_n,l_n}^{({\text d})} \right)\right\}\nonumber\\
&+& \sum\limits_{l_m \in {\cal C}} \left( N_0 \left\| {\bm J}_{l,m} \right\|^2 - 2\ln \gamma_{l_m}^{({\text c})} -1\right) + \sum\limits_{l_n \in {\cal D}} \left[ \left(\gamma_{l_n}^{({\text d})}\right)^2 + N_0 \left|\gamma_{l_n}^{({\text d})} f_{l_n} \right|^2 - 2\ln \gamma_{l_n}^{({\text d})} -1 \right],
\label{F_q_p}
\end{eqnarray}}
\setcounter{equation}{\value{TempEqCnt}}
\end{figure*}
\subsection{Robust Optimization}
\subsubsection{Solving (\ref{optimize_problem3}) for Fixed ${\bm p}$}
For notational convenience, denote
{\setlength\arraycolsep{2pt}
\begin{eqnarray}
&& \gamma_{l_m}^{({\text c})}=\exp \left(\frac{u_{l_m}^{({\text c})}-1}{2}\right), q_{l_m}^{({\text c})} = \sqrt {p_{l_m}^{({\text c})}}, ~\forall~ l_m \in {\cal C}\nonumber\\
&& \gamma_{l_n}^{({\text d})}=\exp \left(\frac{u_{l_n}^{({\text d})}-1}{2}\right), q_{l_n}^{({\text d})} = \sqrt {p_{l_n}^{({\text d})}}, ~\forall~ l_n \in {\cal D}.
\label{gamma_q}
\end{eqnarray}}
\!\!\!Then, for fixed ${\bm p}$, problem (\ref{optimize_problem3}) becomes
{\setlength\arraycolsep{2pt}
\begin{eqnarray}
\!\!\!\mathop {\min }\limits_{{\bm W},{\bm f},{\bm \Gamma},{\bm u}^{({\text d})}} \!\!\!\!\!&&\left\{\sum\limits_{l=1}^L \mathop {\max }\limits_{{\bm \Delta}_l} \sum\limits_{m=1}^M \left[ \left(\gamma_{l_m}^{({\text c})}\right)^2 {\text {MSE}}_{l_m}^{({\text c})} - 2\ln \gamma_{l_m}^{({\text c})} -1 \right]\right.\nonumber\\
&& \left.+ \sum\limits_{l_n \in {\cal D}} \left[ \exp \left(u_{l_n}^{({\text d})} - 1\right) {\text {MSE}}_{l_n}^{({\text d})} - u_{l_n}^{({\text d})} \right] \right\} \nonumber\\
\!\!\!\text{s.t.} \;\;\, && {\text {(\ref{estimation_error_a})}},~ {\text {(\ref{estimation_error_b})}},\nonumber\\
&& \gamma_{l_m}^{({\text c})} > 0, ~\forall~ l_m \in {\cal C},
\label{optimize_problem31}
\end{eqnarray}}
\!\!\!where ${\bm \Gamma}$ is defined as
{\setlength\arraycolsep{2pt}
\begin{eqnarray}
&& \bm \Gamma=\left[ \bm \Gamma_1, \cdots, \bm \Gamma_L\right],\nonumber\\
&& {\bm \Gamma}_l = {\text {diag}} \left(\gamma_{l_1}^{({\text c})},\cdots,\gamma_{l_M}^{({\text c})}\right), ~\forall~ l \in {\cal L}.
\label{Gamma}
\end{eqnarray}}

Since $\bm W_l$, $\bm \Gamma_l$, $f_{l_n}^{({\text d})}$ and $u_{l_n}^{({\text d})}, ~\forall l \!\in\! {\cal L},~ l_n \!\in\! {\cal D}$ are independent, and ${\bm \Delta}_l, ~\forall~ l \in {\cal L}$ have independent uncertainties, problem (\ref{optimize_problem31}) can be divided into $L$ subproblems as (\ref{optimize_problem6}) for any cell $l \in {\cal L}$ and $L\times N$ subproblems as (\ref{optimize_problem25}) for any D2D link $l_n \in {\cal D}$
{\setlength\arraycolsep{2pt}
\begin{subequations}
\begin{align}
\mathop {\min }\limits_{\bm W_l, \bm \Gamma_l}\mathop {\max }\limits_{{\bm \Delta}_l} \quad& \sum\limits_{m=1}^M \left[\left(\gamma_{l_m}^{({\text c})}\right)^2 {\text {MSE}}_{l_m}^{({\text c})} - 2\ln \gamma_{l_m}^{({\text c})} -1 \right]  \label{optimize_problem6_a}\\
\text{s.t.} \quad\quad\; & \left\| {\bm \Delta}_{k_t,l}^{({\text c})} \right\| \leq \varepsilon_{k_t,l}^{({\text c})}, ~\forall~ k_t\in {\cal C},\label{optimize_problem6_b}\\
&\left\| {\bm \Delta}_{k_s,l}^{({\text d})} \right\| \leq \varepsilon_{k_s,l}^{({\text d})}, ~\forall~ k_s \in {\cal D},\label{optimize_problem6_c}\\
&\gamma_{l_m}^{({\text c})} > 0, ~\forall~ l_m \in {\cal C}_l.\label{optimize_problem6_d}
\end{align}
\label{optimize_problem6}
\end{subequations}}

\begin{equation}
\mathop {\min }\limits_{f_{l_n}^{({\text d})}, u_{l_n}^{({\text d})}} \left[\exp \left(u_{l_n}^{({\text d})} - 1\right) {\text {MSE}}_{l_n}^{({\text d})} - u_{l_n}^{({\text d})}\right].
\label{optimize_problem25}
\end{equation}

Since $\bm \delta$ has been determined by (\ref{worst_channel_DU}), problem (\ref{optimize_problem25}) can be optimally solved by respectively obtaining the optimal $f_{l_n}^{({\text d})}$ and $u_{l_n}^{({\text d})}$ based on (\ref{MMSE_filter_DU}) and (\ref{auxiliary_DU3}). In contrast, ${\bm \Delta}_l$ cannot be obtained directly due to the coupling of the cellular channel estimation error vectors. Therefore, it would be more difficult to solve (\ref{optimize_problem6}). In the following theorem, it is shown that problem (\ref{optimize_problem6}) can be cast to an SDP.
\begin{theorem}
For given ${\bm p}$, problem (\ref{optimize_problem6}) can be optimized over $\bm W_l$ and $\bm \Gamma_l$ efficiently as an SDP.
\label{theorem_1}
\end{theorem}
\itshape \textbf{Proof:}  \upshape
See Appendix B.
\hfill $\Box$

\subsubsection{Solving (\ref{optimize_problem3}) for Fixed ${\bm W}$, ${\bm f}$ and ${\bm u}$}
When ${\bm W}$, ${\bm f}$ and ${\bm u}$ have been determined, based on the definition of MSE, the objective function of (\ref{optimize_problem3}) can be rewritten as (\ref{F_q_p}), shown as the bottom of the next page, where ${\bm J}_l$, ${\bm J}_{l,m}$ and $\bm e_m$ are defined in (\ref{MSE_CU}). From (\ref{F_q_p}), it is seen that $q_{l_m}^{({\text c})}, ~\forall~ l_m \in {\cal C}$ can be obtained by separately solving the following subproblems
\setcounter{equation}{38}
{\setlength\arraycolsep{2pt}
\begin{subequations}
\begin{align}
\mathop {\min }\limits_{q_{l_m}^{({\text c})}}\mathop {\max }\limits_{{\bm \Delta}_{l_m}^{({\text c})}} \quad& \left\|q_{l_m}^{({\text c})} {\bm J}_l^H {\bm h}_{l_m,l}^{({\text c})} \!-\! \gamma_{l_m}^{({\text c})} {\bm e}_m\right\|^2 \!+\! \sum\limits_{k\neq l} \left\| q_{l_m}^{({\text c})} {\bm J}_k^H {\bm h}_{l_m,k}^{({\text c})}\right\|^2\nonumber\\
& + \left(q_{l_m}^{({\text c})}\right)^2 \sum\limits_{k_s \in {\cal D}} \left|\gamma_{k_s}^{({\text d})} f_{k_s}^H {\bar g}_{l_m,k_s}^{({\text c})} \right|^2 \label{optimize_problem8_a}\\
\text{s.t.} \quad\quad\; & 0 \leq q_{l_m}^{({\text c})} \leq \sqrt{P_{l_m}^{({\text c})}},\label{optimize_problem8_b}\\
&\left\| {\bm \Delta}_{l_m,k}^{({\text c})} \right\| \leq \varepsilon_{l_m,k}^{({\text c})}, ~\forall~ k\in {\cal L},\label{optimize_problem8_c}
\end{align}
\label{optimize_problem8}
\end{subequations}}
\!\!\!\!\!\!where ${\bm \Delta}_{l_m}^{({\text c})} = \left[ {\bm \Delta}_{l_m,1}^{({\text c})}, \cdots, {\bm \Delta}_{l_m,L}^{({\text c})}\right]$.

Due to the maximum interference constraints (\ref{optimize_problem_d}), ${\bm q}^{({\text d})}$ cannot be obtained by separately finding $q_{l_n}^{({\text d})}, ~\forall~ l_n \in {\cal D}$. Therefore, the following problem is solved to obtain the optimal ${\bm q}^{({\text d})}$

{\setlength\arraycolsep{2pt}
\begin{subequations}
\begin{align}
\mathop {\min }\limits_{{\bm q}^{({\text d})}} \mathop {\max }\limits_{{\bm \Delta}^{({\text d})}} \quad& \sum\limits_{l_n \in {\cal D}} \left\{\sum\limits_{k=1}^L \left\|q_{l_n}^{({\text d})} {\bm J}_k^H {\bm h}_{l_n,k}^{({\text d})} \right\|^2 \right. \nonumber\\
&\left. + \theta_{l_n} \left(q_{l_n}^{({\text d})}\right)^2 - 2 \phi_{l_n} q_{l_n}^{({\text d})}\right\} \label{optimize_problem5_a}\\
\text{s.t.} \quad\quad\; & 0 \leq q_{l_n}^{({\text d})} \leq \sqrt{P_{l_n}^{({\text d})}}, ~\forall~ l_n \in {\cal D}, \label{optimize_problem5_b}\\
& \sum\limits_{k_s\in {\cal D}} \rho_{k_s,l} \left(q_{k_s}^{({\text d})}\right)^2 \leq a_l, ~\forall~ l \in {\cal L}, \label{optimize_problem5_c}\\
&\left\| {\bm \Delta}_{l_n,k}^{({\text d})} \right\| \leq \varepsilon_{l_n,k}^{({\text d})}, ~\forall~ l_n \in {\cal D},~ k\in {\cal L},\label{optimize_problem5_d}
\end{align}
\label{optimize_problem5}
\end{subequations}}
\!\!\!where
{\setlength\arraycolsep{2pt}
\begin{eqnarray}
{\bm q}^{({\text d})} &= & \left(q_{1_1}^{({\text d})}, \cdots, q_{1_N}^{({\text d})}, \cdots, q_{L_N}^{({\text d})}\right)^T, \nonumber\\
{\bm \Delta}^{({\text d})} &= & \left[ {\bm \Delta}_{1_1,1}^{({\text d})}, \cdots, {\bm \Delta}_{1_1,L}^{({\text d})}, \cdots, {\bm \Delta}_{L_N,L}^{({\text d})}\right], \nonumber\\
\theta_{l_n}^{({\text d})} &= & \sum\limits_{k_s \in {\cal D}} \left|\gamma_{k_s}^{({\text d})} f_{k_s}^H {\bar g}_{l_n,k_s}^{({\text d})} \right|^2, ~\forall~ l_n \in {\cal D},\nonumber\\
\phi_{l_n}^{({\text d})} &= & \left(\gamma_{l_n}^{({\text d})}\right)^2 {\text {Re}}\left( f_{l_n}^H {\bar g}_{l_n,l_n}^{({\text d})} \right), ~\forall~ l_n \in {\cal D}.
\label{parameters}
\end{eqnarray}}

To solve the robust optimization problems (\ref{optimize_problem8}) and (\ref{optimize_problem5}), the following theorem is given, which shows that they have computationally efficient solutions.
\begin{theorem}
For given ${\bm W}$, ${\bm f}$ and ${\bm u}$, either problem (\ref{optimize_problem8}) or problem (\ref{optimize_problem5}) can be posed as an SDP.
\label{theorem_2}
\end{theorem}
\itshape \textbf{Proof:}  \upshape
See Appendix C.
\hfill $\Box$

Based on the above analysis, problem (\ref{optimize_problem3}) can be solved by iteratively optimizing ${\bm W}$, ${\bm f}$, ${\bm u}$ and ${\bm p}$. The detailed steps are summarized in Algorithm \ref{algorithm1}.
\begin{algorithm}[h]
\begin{algorithmic}[1]
\caption{Robust Transmission Design (RTD)}
\label{RTD}
\State Set $j=0$, initialize $\bm p(j)$. $q_{l_m}^{({\text c})}(j)=\sqrt{p_{l_m}^{({\text c})}(j)}, ~\forall~ l_m \in {\cal C}$, $q_{l_n}^{({\text d})}(j)=\sqrt{p_{l_n}^{({\text d})}(j)}, ~\forall~ l_n \in {\cal D}$.
\Repeat
\State Obtain $\bm J_l(j+1)$ and $\bm \Gamma_l(j+1), ~\forall~ l \in {\cal L}$ by solving SDP (\ref{optimize_problem10}).
\vspace{0.1em}
\State $\bm W_l(j+1) = \bm J_l(j+1)\bm \Gamma_l(j+1)^{-1}$, $u_{l_m}^{({\text c})}(j+1)=1+2\ln \gamma_{l_m}^{({\text c})}(j+1), ~\forall~ l \in {\cal L},~ l_m \in {\cal C}_l$.
\vspace{0.1em}
\State Obtain $f_{l_n}^{({\text d})}(j+1)$ and $u_{l_n}^{({\text d})}(j+1), ~\forall~ l_n \in {\cal D}$ directly from (\ref{MMSE_filter_DU}) and (\ref{auxiliary_DU3}).
\vspace{0.1em}
\State Obtain $q_{l_m}^{({\text c})}(j+1), ~\forall~ l_m \in {\cal C}$ and $\bm q^{({\text d})}(j+1)$ by solving SDP (\ref{optimize_problem16}) and (\ref{optimize_problem21}), respectively.
\vspace{0.1em}
\State $p_{l_m}^{({\text c})}(j+1)=\left[q_{l_m}^{({\text c})}(j+1)\right]^2, ~\forall~ l_m \in {\cal C}$, $p_{l_n}^{({\text d})}(j+1)=\left[q_{l_n}^{({\text d})}(j+1)\right]^2, ~\forall~ l_n \in {\cal D}$.
\vspace{0.1em}
\State Let $j=j+1$.
\Until convergence
\label{algorithm1}
\end{algorithmic}
\end{algorithm}
\subsection{Convergence and Complexity Analysis}
Since an alternative algorithm is proposed to solve (\ref{optimize_problem3}), it is necessary to characterize the convergence behavior of Algorithm \ref{algorithm1}.
\begin{theorem}
The iterative robust transmission design given in Algorithm \ref{algorithm1} converges to a suboptimal solution of problem (\ref{optimize_problem3}).
\label{theorem_3}
\end{theorem}
\itshape \textbf{Proof:}  \upshape
See Appendix D.
\hfill $\Box$

Besides the convergence behavior, it is also necessary to analyze the computational complexity of the proposed RTD algorithm. In the following, order notation ${\cal O}(\cdot)$ is adopted to characterize the computational complexity of Algorithm \ref{algorithm1}. According to \cite{vandenberghe1996semidefinite, hanif2014computationally, luo2004transceiver}, solving an SDP involves a complexity of ${\cal O}(i^{6.5} \log \lambda)$ for accuracy $\lambda$, where $i$ denotes the dimension of matrix variables. Since ${\bm F}_{k_t,l}^{({\text c})}, {\bm F}_{k_s,l}^{({\text d})}, ~\forall~ l \in {\cal L},~ k_t\in {\cal C}, ~\forall~ k_s\in {\cal D}$ are all $(1+M+B)\times(1+M+B)$ dimensional matrices, the complexity of solving SDP (\ref{optimize_problem10}) is ${\cal O}([L(M+N)(1+M+B)]^{6.5} \log \lambda)$. Similarly, obtaining $\bm q$ by solving SDP (\ref{optimize_problem16}) and (\ref{optimize_problem21}) also requires complexity of ${\cal O}([L(M+N)(1+M+B)]^{6.5} \log \lambda)$. Let $I_{\text {ite}}$ denote the iteration numbers of the proposed RTD algorithm, then, the total complexity for solving (\ref{optimize_problem3}) is ${\cal O}(I_{\text {ite}} [L(M+N)(1+M+B)]^{6.5} \log \lambda)$. It is observed from
simulations that the RTD algorithm converges after a few iterations, so the complexity is low and acceptable.
\section{Simulation Results}
\label{Simulation_Results}
\begin{figure}
  \centering
  \includegraphics[scale=0.5]{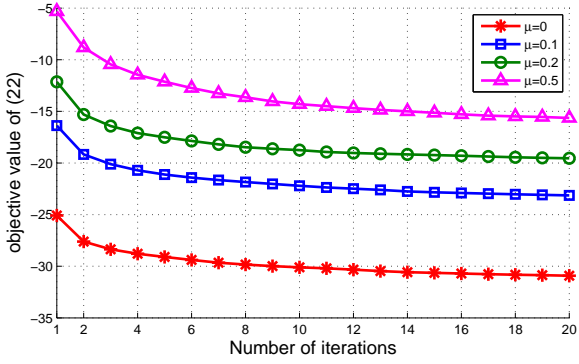}
  \caption{Convergence behaviors of the proposed RTD algorithm with $L=2$, $M=2$, $N=3$, $B=4$, $P=20$ dBm, $a=-80$ dBm  and $D_{\text{max}}=100$.}
  \label{Fig.2}
\end{figure}

In this section, simulation results are presented to evaluate the performance of the proposed RTD algorithm. Based on Theorem \ref{theorem_1}, Theorem \ref{theorem_2} and Algorithm \ref{algorithm1}, it is required to solve SDPs (\ref{optimize_problem10}), (\ref{optimize_problem16}) and (\ref{optimize_problem21}) to obtain a suboptimal solution of (\ref{optimize_problem3}). CVX, a toolbox developed in MATLAB for solving convex problems \cite{grant2008cvx}, is used to solve SDPs. All simulation results are obtained by averaging over 1000 channel realizations, and each channel realization is obtained by generating a random user distribution as well as a random set of fading coefficients.

Consider a multi-cell D2D underlaid cellular system. All users are distributed uniformly and it is assumed that no user is closer to a BS than 20 meters. The distance between a D2D-Tx and its associated receiver is uniformly distributed in the range of $[0\,\text{m}, D_{\text{max}}\,\text{m}]$. The pathloss exponent and the standard deviation of log-normal shadowing fading are respectively set to be 3.7 and 8 dB \cite{access2010further}. The noise power is $N_0 = -100$ dBm. For brevity, equal maximum interference threshold at all BSs and equal maximum transmit power for all mobile transmitters are assumed, i.e., $a_l = a, ~\forall~ l \in {\cal L}$ and $P_{l_m}^{({\text c})}=P_{l_n}^{({\text d})}=P, ~\forall~ l_m \in {\cal C},~ l_n \in {\cal D}$. Since this paper adopts the bounded CSI error model to characterize CSI impairment and quantization errors are the main source of CSI uncertainty for this model, a channel estimation error vector would be closely related to the corresponding channel vector estimate. Therefore, it is assumed that $\varepsilon_{k_t,l}^{({\text c})}= \mu \left\| {\tilde {\bm h}}_{k_t,l}^{({\text c})}\right\|$, $\varepsilon_{k_s,l}^{({\text d})}= \mu \left\| {\tilde {\bm h}}_{k_s,l}^{({\text d})}\right\|$, $\epsilon_{k_t,l_n}^{({\text c})}= \mu \left| {\tilde g}_{k_t,l_n}^{({\text c})}\right|$ and $\epsilon_{k_s,l_n}^{({\text d})}= \mu \left| {\tilde g}_{k_s,l_n}^{({\text d})}\right|$, $ \forall~ l \in {\cal L},~ k_t\in {\cal C},~ l_n,~ k_s\in {\cal D}$, where $\mu \in [0, 1)$ is a metric used for evaluating the CSI error level \cite{xu2015robust}.
\subsection{Convergence Behaviors of the Proposed RTD Algorithm}
\begin{figure}
  \centering
  \includegraphics[scale=0.5]{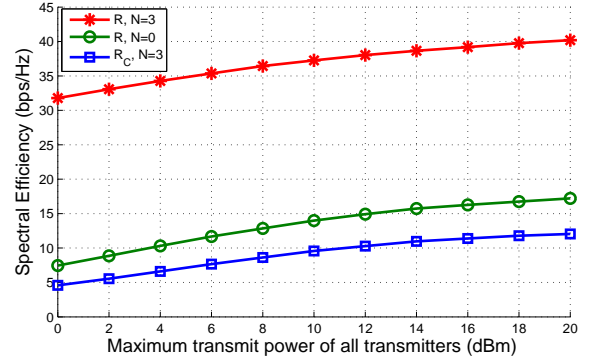}
  \caption{Sum SE comparison between D2D underlaid cellular system and the conventional cellular system with $L=2$, $M=2$, $B=4$, $a=-80$ dBm, $\mu=0.3$ and $D_{\text{max}}=100$.}
  \label{Fig.3}
\end{figure}
Based on the analysis in Section \ref{Problem_Analysis_and_Robust_Optimization}, the original worst-case SE maximization problem (\ref{optimize_problem}) is first transformed into (\ref{optimize_problem26}) using Theorem \ref{theorem_0}. By exchanging the positions of $\max$ and $\min$, an upper bound on the objective function of problem (\ref{optimize_problem26}) is obtained and (\ref{optimize_problem26}) is further simplified as (\ref{optimize_problem2}). Then, the proposed RTD algorithm is adopted to obtain a suboptimal solution of (\ref{optimize_problem2}). Fig. \ref{Fig.2} depicts the convergence behaviors of the proposed RTD algorithm under different values of channel error level $\mu$. It can be seen from this figure that the objective value of (\ref{optimize_problem2}) monotonically decreases during the iterative procedure and converges well in about 20 iterations for all considered configurations. Moreover, Fig. \ref{Fig.2} also shows that the objective value of (\ref{optimize_problem2}) grows with respect to (w.r.t.) $\mu$. This is consistent with intuition since a larger $\mu$ usually results in more severe CSI uncertainties and, hence, increases the MSE of all links. Note that when $\mu=0$, i.e., ${\bm \Delta}={\bm 0}$ and ${\bm \delta}={\bm 0}$, problem (\ref{optimize_problem2}) becomes (\ref{optimize_problem26}). In this case, a suboptimal solution of the original problem (\ref{optimize_problem}) can be obtained by using the proposed RTD algorithm.

\subsection{Comparison to the Conventional Cellular Communication}
To characterize the effect of D2D communication to the conventional cellular system (i.e., $N=0$), the sum SE gains of the system and the performance loss of CUs resulted from D2D communication are investigated. Let $R$ and $R_{\text C}$ denote the sum SE of all links and the sum SE of all CUs, respectively.

Fig. \ref{Fig.3} depicts the sum SE versus the maximum power of mobile transmitters. As expected, both $R$ and $R_{\text C}$ increase w.r.t. the maximum transmit power $P$. Fig. \ref{Fig.3} also shows that D2D underlaid communication can provide significant performance gains over the conventional cellular communication. Specifically, for the case with $P=10$ dBm, the sum SE of the conventional cellular system is increased by over 150\% by D2D communication with $N=3$. On the other hand, the performance of CUs may be affected by underlaid D2D communication due to cochannel interference. Since it is assumed in this paper that the interference signal from all D2D-Txs to each BS is upper bounded by $a$, the performance loss of CUs is much smaller than the performance gains brought by D2D communication, which can be verified by Fig. \ref{Fig.3}.

\begin{figure}
  \centering
  \includegraphics[scale=0.5]{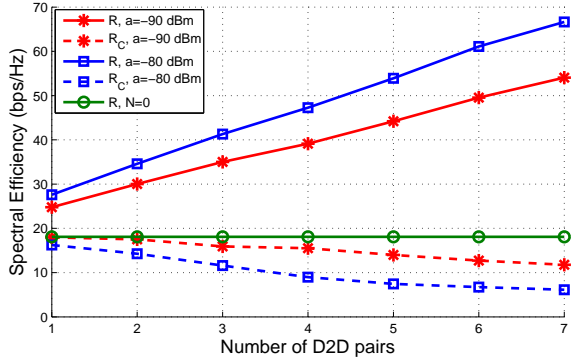}
  \caption{Sum SE of all CUs and Sum SE of the system versus $N$ with $L=2$, $M=2$, $B=4$, $P=20$ dBm, $\mu=0.5$ and $D_{\text{max}}=100$.}
  \label{Fig.4}
\end{figure}

\begin{figure}
  \centering
  \includegraphics[scale=0.5]{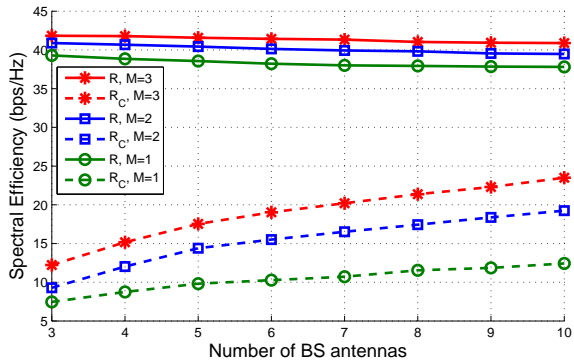}
  \caption{Sum SE of all CUs and Sum SE of the system versus $B$ with $L=2$, $N=3$, $P=20$ dBm, $a=-80$ dBm, $\mu=0.5$ and $D_{\text{max}}=100$.}
  \label{Fig.5}
\end{figure}

Fig. \ref{Fig.4} plots the sum SE of all CUs and sum SE of the system versus the number of D2D pairs under different values of $a$. As a benchmark, the sum SE of the system without D2D communication is also depicted. From this figure, it is observed that the sum SE of the system, $R$, increases almost linearly with $N$, whereas $R_{\text C}$ decreases with $N$ due to cochannel interference resulted from D2D communication. When $a=-90$ dBm, the performance loss of CUs is small. For $a=-80$ dBm with small $N$, the performance loss of CUs appears also marginal, while as $N$ grows larger than 4, noticeable cellular SE loss can be observed. This is because when aiming to maximize $R$, though the power of the interference signal from all D2D-Txs to each BS is upper bounded by $a$, the transmit power of CUs may be suppressed to increase the SE of DUs since DUs usually possess better channel conditions than CUs due to short transmission distance, especially when $N$ is large. In this case, one can always reduce the performance loss of CUs by decreasing $a$ to further protect cellular communication. Combining Fig. \ref{Fig.3} and Fig. \ref{Fig.4}, one can conclude that it is an effective way to increase the system throughput by introducing D2D communication to a conventional cellular communication system while causing tolerable interference to CUs.

\begin{figure}
  \centering
  \includegraphics[scale=0.5]{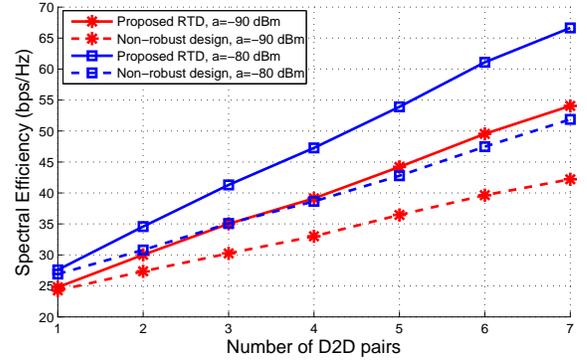}
  \caption{Sum SE of the system versus $N$ with $L=2$, $M=2$, $B=4$, $P=20$ dBm, $\mu=0.5$ and $D_{\text{max}}=100$.}
  \label{Fig.6}
\end{figure}

In Fig. \ref{Fig.5}, the effects of the number of BS antennas and the number of CUs are investigated. From this figure it can be found that both $R$ and $R_{\text C}$ increase with $M$, which is consistent with intuition. However, the increasing range of $R$ is smaller than that of $R_{\text C}$. This is because as $M$ grows, the co-channel interference experienced by DUs increases, leading to the reduction of D2D communication throughput. In addition, Fig. \ref{Fig.5} also shows that for all considered cases, as $B$ grows, $R_{\text C}$ prominently increases, while $R$ experiences a very slight decline. This can be explained by constraints (\ref{optimize_problem_d}). From (\ref{SINR_reduce_strict}) and the assumption that $\varepsilon_{k_s,l}^{({\text d})}= \mu \left\| {\tilde {\bm h}}_{k_s,l}^{({\text d})}\right\|$, it follows that $\rho_{k_s,l}=(1+\mu)^2 \left\| {\tilde {\bm h}}_{k_s,l}^{({\text d})}\right\|^2$. Since the elements in ${\tilde {\bm h}}_{k_s,l}^{({\text d})}$ are independent and follow the same distribution, as $B$ grows, stronger constraints are imposed to DUs, i.e., DUs have to transmit in a relatively low power. Therefore, the sum SE of DUs decreases with $B$. In this case, one can always increase $R$ by increasing $a$ to relax constraints (\ref{optimize_problem_d}).

\subsection{Performance of the Proposed RTD Algorithm}
For a single-cell single-input single-output (SISO) network, the problem of maximizing the sum system SE under perfect CSI was studied in \cite{zhao2015resource}. However, for a multi-cell MIMO system with several CUs and several D2D pairs in each cell, as stated in Section \ref{Introduction}, the considered RTD problem under CSI impairment has not been studied. Therefore, to evaluate the performance of the proposed RTD algorithm, as in \cite{tajer2011robust, xu2015robust}, the simulation results obtained by using the `non-robust' design are used as the benchmark, and compared with that obtained by the proposed RTD algorithm. In particular, the non-robust scheme tries to jointly optimize $\bm p$ and $\bm W$ by simply treating the estimated channel coefficients as the true channel, and the non-robust benchmark is obtained by using Algorithm \ref{algorithm1} with $\mu$ set to be 0.

\begin{figure}
  \centering
  \includegraphics[scale=0.5]{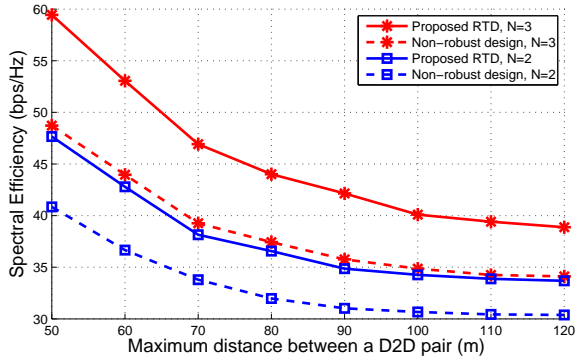}
  \caption{Sum SE of the system versus $D_{\text{max}}$ with $L=2$, $M=2$, $B=4$, $P=20$ dBm, $a=-80$ dBm and $\mu=0.5$.}
  \label{Fig.7}
\end{figure}

\begin{figure}
  \centering
  \includegraphics[scale=0.5]{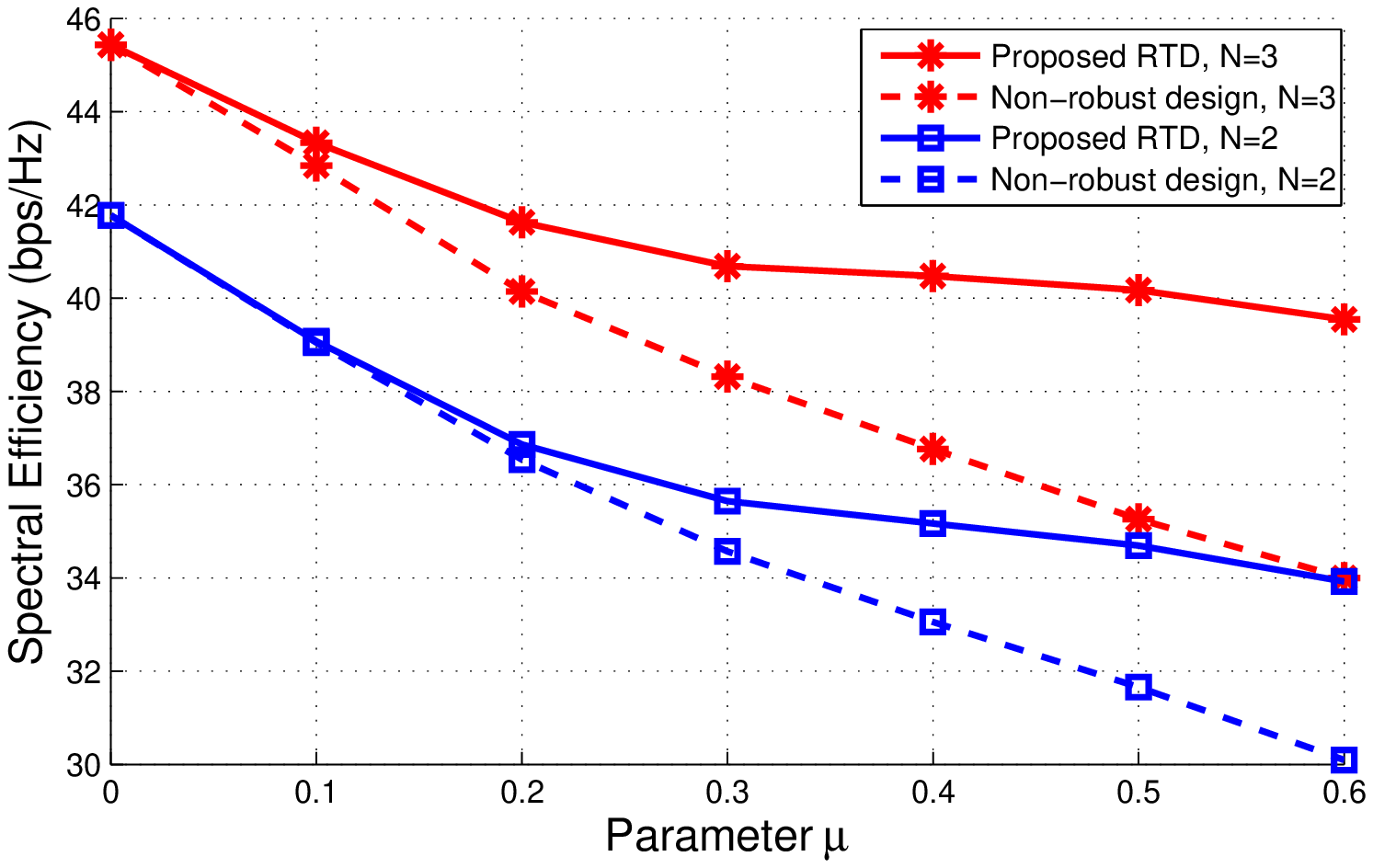}
  \caption{Sum SE of the system versus $\mu$ with $L=2$, $M=2$, $B=4$, $P=20$ dBm, $a=-80$ dBm and $D_{\text{max}}=100$.}
  \label{Fig.8}
\end{figure}

Fig. \ref{Fig.6} compares the proposed RTD algorithm with the non-robust design versus the number of D2D pairs under different values of $a$. It can be seen that in the small $N$ regime, compared with the non-robust design, a small SE gain is obtained by Algorithm \ref{algorithm1}, while as $N$ grows large, the SE gap increases greatly. In particular, when $N=5$, compared with the non-robust design scheme, the sum SE of the system can be respectively increased by about 25\% and 20\% for the cases with $a=-80$ dBm and $a=-90$ dBm via adopting the proposed RTD algorithm. The effect of the maximum distance between a D2D pair is investigated in Fig. \ref{Fig.7}. As expected, the sum SE of the system decreases with $D_{\text{max}}$ for all considered cases.

\begin{figure}
  \centering
  \includegraphics[scale=0.5]{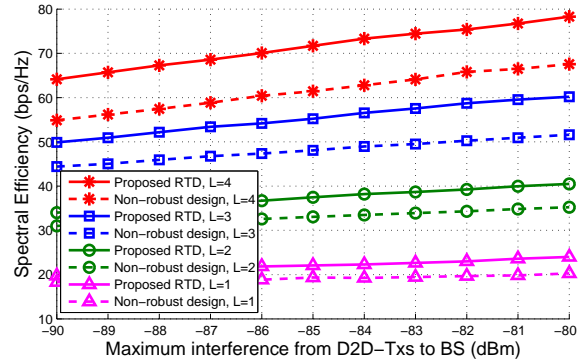}
  \caption{Sum SE of the system versus $a$ with $M=2$, $N=3$, $B=4$, $P=20$ dBm and $D_{\text{max}}=100$.}
  \label{Fig.9}
\end{figure}

Fig. \ref{Fig.8} depicts the sum SE comparison versus channel error level $\mu$ under different values of $N$. It shows that the sum SE of the system monotonically decreases with $\mu$, which is consistent with intuition. In addition, it can also be seen that the performance gains brought by Algorithm \ref{algorithm1} over the non-robust design increase prominently with $\mu$, which indicates that Algorithm \ref{algorithm1} is more efficient in obtaining a higher throughput of the system when channel estimation suffers from serious uncertainties. The effect of the maximum interference from all D2D-Txs to each BS is investigated in Fig. \ref{Fig.9} for different network sizes. As expected, the sum SE of the system increases with $a$ and the network size. It also shows that in contrast to the non-robust design scheme, the sum SE gains brought by the proposed RTD algorithm grows with $L$. Hence, Algorithm \ref{algorithm1} is suitable for networks with different sizes.

\section{Conclusions}
\label{Conclusions}
In this paper, the robust transmission design for a multi-cell D2D underlaid cellular system when BSs only have imperfect CSI of all links has been studied. To account for CSI uncertainties, this paper aims to maximize the worst-case sum rate of the system while guaranteeing that the interference signal from all D2D-Txs to each BS is power-limited. To solve the nonconvex problem, it is first transformed into a more tractable form by replacing the objective function of the original problem with its lower bound. Then, the resulted problem is decomposed into several convex SDP subproblems, and an iterative algorithm is proposed to obtain a suboptimal solution. Simulation results show that the performance of the conventional cellular systems can be significantly improved by D2D communication while endurable interference is caused to CUs. In addition, the proposed robust transmission design algorithm greatly outperforms the non-robust transmission design in terms of system SE.
\appendices
\section{Proof of Theorem \ref{theorem_0}}
\label{Appendix_A}
From (\ref{Received_signal_BS}), the MMSE receive filter at BS $l$ for detecting $x_{l_m}^{({\text c})}$ is given by \footnote{From the expressions of ${\bm G}_{l_m}$ and ${\bm A}_{l_m}$, it is known that ${\bm G}_{l_m}$ and ${\bm A}_{l_m}+ {\bm G}_{l_m}$ are both positive definite matrices, and are thus reversible.}
\setcounter{equation}{0}
\renewcommand{\theequation}{\thesection.\arabic{equation}}
{\setlength\arraycolsep{2pt}
\begin{eqnarray}
{\bm w}_{l_m}^{\text {MMSE}} &=& \arg \mathop {\min }\limits_{{\bm w}_{l_m}} {\mathbb E} \left\{ \left|{\bm w}_{l_m}^H {\bm y}_l^{({\text c})}- {x}_{l_m}^{({\text c})} \right|^2\right\} \nonumber\\
&=& \sqrt{p_{l_m}^{({\text c})}} \left( {\bm A}_{l_m} + {\bm G}_{l_m}\right)^{-1} {\bm h}_{l_m,l}^{({\text c})},
\label{MMSE_filter}
\end{eqnarray}}
\!\!where ${\bm A}_{l_m}= p_{l_m}^{({\text c})} {\bm h}_{l_m,l}^{({\text c})} \left( {\bm h}_{l_m,l}^{({\text c})} \right)^H$. Using (\ref{MMSE_filter}), the MMSE of cellular link $l_m$ can be written as
{\setlength\arraycolsep{2pt}
\begin{eqnarray}
&&{\text {MMSE}}_{l_m}^{({\text c})}= {\mathbb E} \left\{ \left|\left( {\bm w}_{l_m}^{\text {MMSE}} \right)^H {\bm y}_l^{({\text c})}- { x}_{l_m}^{({\text c})} \right|^2\right\} \nonumber\\
&&= \left( {\bm w}_{l_m}^{\text {MMSE}} \right)^H \left( {\bm A}_{l_m} + {\bm G}_{l_m}\right) {\bm w}_{l_m}^{\text {MMSE}} +1 \nonumber\\
&& \quad - \sqrt{p_{l_m}^{({\text c})}} \left( {\bm w}_{l_m}^{\text {MMSE}} \right)^H {\bm h}_{l_m,l}^{({\text c})} - \sqrt{p_{l_m}^{({\text c})}} \left( {\bm h}_{l_m,l}^{({\text c})} \right)^H {\bm w}_{l_m}^{\text {MMSE}}\nonumber\\
&&= 1 - p_{l_m}^{({\text c})} \left( {\bm h}_{l_m,l}^{({\text c})} \right)^H \left( {\bm A}_{l_m} + {\bm G}_{l_m}\right)^{-1} {\bm h}_{l_m,l}^{({\text c})}\nonumber\\
&&= \frac{1}{1+p_{l_m}^{({\text c})} \left( {\bm h}_{l_m,l}^{({\text c})} \right)^H  {\bm G}_{l_m}^{-1} {\bm h}_{l_m,l}^{({\text c})}}.
\label{MMSE_CU}
\end{eqnarray}}
\!\!\!To prove equation (\ref{MMSE_SINR_CU}), in the following, it is proven that the post-processing SINR of CU $l_m$ satisfies $\text {SINR}_{l_m}^{({\text c})}=p_{l_m}^{({\text c})} \left( {\bm h}_{l_m,l}^{({\text c})} \right)^H  {\bm G}_{l_m}^{-1} {\bm h}_{l_m,l}^{({\text c})}$ by checking the following term
{\setlength\arraycolsep{2pt}
\begin{eqnarray}
&&\quad\frac{\text {SINR}_{l_m}^{({\text c})}}{p_{l_m}^{({\text c})} \left( {\bm h}_{l_m,l}^{({\text c})} \right)^H  {\bm G}_{l_m}^{-1} {\bm h}_{l_m,l}^{({\text c})}}
 \nonumber\\
&&\mathop = \limits^{(\text{a})} \frac{\left( {\bm w}_{l_m}^{\text {MMSE}} \right)^H  {\bm A}_{l_m} {\bm w}_{l_m}^{\text {MMSE}}}{\left( {\bm w}_{l_m}^{\text {MMSE}} \right)^H  {\bm G}_{l_m} {\bm w}_{l_m}^{\text {MMSE}} p_{l_m}^{({\text c})} \left( {\bm h}_{l_m,l}^{({\text c})} \right)^H  {\bm G}_{l_m}^{-1} {\bm h}_{l_m,l}^{({\text c})}}\nonumber\\
&&\mathop = \limits^{(\text{b})} \frac{\left( {\bm w}_{l_m}^{\text {MMSE}} \right)^H  {\bm A}_{l_m} {\bm w}_{l_m}^{\text {MMSE}}}{\left( {\bm w}_{l_m}^{\text {MMSE}} \right)^H  {\bm G}_{l_m} \sqrt{p_{l_m}^{({\text c})}} \left( {\bm A}_{l_m} + {\bm G}_{l_m}\right)^{-1} {\bm A}_{l_m} {\bm G}_{l_m}^{-1} {\bm h}_{l_m,l}^{({\text c})}},\nonumber\\
\label{MMSE_temp}
\end{eqnarray}}
\!\!\!where (a) uses (\ref{SINR_CU}) and (b) follows by substituting (\ref{MMSE_filter}). Since
{\setlength\arraycolsep{2pt}
\begin{eqnarray}
&&{\bm G}_{l_m} \sqrt{p_{l_m}^{({\text c})}} \left( {\bm A}_{l_m} + {\bm G}_{l_m}\right)^{-1} {\bm A}_{l_m} {\bm G}_{l_m}^{-1} {\bm h}_{l_m,l}^{({\text c})}\nonumber\\
&=& {\bm G}_{l_m} \sqrt{p_{l_m}^{({\text c})}} \left( {\bm A}_{l_m} \!+\! {\bm G}_{l_m}\right)^{-1} \left({\bm A}_{l_m} \!+\! {\bm G}_{l_m}\!-\! {\bm G}_{l_m} \right){\bm G}_{l_m}^{-1} {\bm h}_{l_m,l}^{({\text c})}\nonumber\\
&=&\sqrt{p_{l_m}^{({\text c})}} \left( {\bm A}_{l_m} + {\bm G}_{l_m}\right) \left( {\bm A}_{l_m} + {\bm G}_{l_m}\right)^{-1} {\bm h}_{l_m,l}^{({\text c})}\nonumber\\
&& - \sqrt{p_{l_m}^{({\text c})}} {\bm G}_{l_m} \left( {\bm A}_{l_m} + {\bm G}_{l_m}\right)^{-1} {\bm h}_{l_m,l}^{({\text c})}\nonumber\\
&=&\sqrt{p_{l_m}^{({\text c})}} {\bm A}_{l_m} \left( {\bm A}_{l_m} + {\bm G}_{l_m}\right)^{-1} {\bm h}_{l_m,l}^{({\text c})}\nonumber\\
&=& {\bm A}_{l_m} {\bm w}_{l_m}^{\text {MMSE}},
\label{MMSE_temp1}
\end{eqnarray}}
\!\!\!combining (\ref{MMSE_CU}), (\ref{MMSE_temp}) and (\ref{MMSE_temp1}), it can be concluded that
\begin{equation}
{\text {MMSE}}_{l_m}^{({\text c})}=\frac{1}{1+\text {SINR}_{l_m}^{({\text c})}}.
\end{equation}

Similarly, as for D2D link $l_n$, in the worst-case, the MMSE receive filter at D2D-Rx $l_n$ is given by
{\setlength\arraycolsep{2pt}
\begin{eqnarray}
&&f_{l_n}^{\text {MMSE}} = \arg \mathop {\min }\limits_{f_{l_n}} {\mathbb E} \left\{ \left|f_{l_n}^H {\bar y}_{l_n}^{({\text d})}- {x}_{l_n}^{({\text d})} \right|^2\right\} \nonumber\\
&&= \frac{\sqrt{p_{l_n}^{({\text d})}} {\bar g}_{l_n,l_n}^{({\text d})}}{\sum\limits_{k_t \in {\cal C}} p_{k_t}^{({\text c})} \left|{\bar g}_{k_t,l_n}^{({\text c})}\right|^2 + \sum\limits_{k_s \in {\cal D}} p_{k_s}^{({\text d})} \left|{\bar g}_{k_s,l_n}^{({\text d})}\right|^2 + N_0},
\label{MMSE_filter_DU}
\end{eqnarray}}
\!\!where ${\bar y}_{l_n}^{({\text d})}$ is obtained by replacing the channel coefficients in (\ref{Received_signal_DU}) with (\ref{worst_channel_DU}), i.e.,
\begin{equation}
{\bar y}_{l_n}^{({\text d})} = \!\!\sum\limits_{k_t \in {\cal C}} \!\!\sqrt {p_{k_t}^{({\text c})}} {\bar g}_{k_t,l_n}^{({\text c})} x_{k_t}^{({\text c})} + \!\!\sum\limits_{k_s \in {\cal D}}\!\! \sqrt {p_{k_s}^{({\text d})}} {\bar g}_{k_s,l_n}^{({\text d})} x_{k_s}^{({\text d})} + z_{l_n}^{({\text d})}.
\label{Received_signal_DU1}
\end{equation}
Then, the worst-case ${\text {MMSE}}_{l_n}^{({\text d})}$ can be written as
{\setlength\arraycolsep{2pt}
\begin{eqnarray}
&&{\text {MMSE}}_{l_n}^{({\text d})}= {\mathbb E} \left\{ \left|\left(f_{l_n}^{\text {MMSE}}\right)^H {\bar y}_{l_n}^{({\text d})}- {x}_{l_n}^{({\text d})} \right|^2\right\} \nonumber\\
&&= 1- \frac{p_{l_n}^{({\text d})} \left|{\bar g}_{l_n,l_n}^{({\text d})}\right|^2}{\sum\limits_{k_t \in {\cal C}} p_{k_t}^{({\text c})} \left|{\bar g}_{k_t,l_n}^{({\text c})}\right|^2 + \sum\limits_{k_s \in {\cal D}} p_{k_s}^{({\text d})} \left|{\bar g}_{k_s,l_n}^{({\text d})}\right|^2 + N_0}.\quad
\label{MMSE_DU}
\end{eqnarray}}
\!\!Combining (\ref{worst_SINR_DU}) and (\ref{MMSE_DU}), it can be easily seen that
\begin{equation}
{\text {MMSE}}_{l_n}^{({\text d})}=  \frac{1}{1+\text {sinr}_{l_n}^{({\text d})}}.
\label{MMSE_DU1}
\end{equation}

Thus, Theorem \ref{theorem_0} is proven.
\section{Proof of Theorem \ref{theorem_1}}
\label{Appendix_B}
Before proving Theorem \ref{theorem_1}, the following useful result from \cite{boyd1994linear, ben2003extended} is first given.
\begin{lemma}
For any given matrices ${\bm X}$, ${\bm Y}$ and ${\bm Z}$, and a Hermitian matrix ${\bm \Lambda} = {\bm \Lambda}^H$, the inequality
\setcounter{equation}{0}
\begin{equation}
{\bm \Lambda} \succeq {\bm X}^H {\bm Z} {\bm Y} + {\bm Y}^H {\bm Z}^H {\bm X}, ~\forall~ {\bm Z}:~ \left\| {\bm Z}\right\|\leq \varepsilon
\end{equation}
holds if and only if
\begin{equation}
\exists~ \eta\geq 0 \,{\text {such that}} \,\left[\! \begin{array}{l} {\bm \Lambda}-\eta {\bm X}^H {\bm X}  \;\;-\varepsilon{\bm Y}^H\\ \quad\;\;\;-\varepsilon{\bm Y}\quad\quad\quad \eta{\bm I} \end{array} \!\right]\succeq 0.
\end{equation}
\label{lemma_1}
\end{lemma}

Combining (\ref{Received_signal_BS}) and the definition of MSE, the objective function of problem (\ref{optimize_problem6}) can be rewritten as
{\setlength\arraycolsep{2pt}
\begin{eqnarray}
&& \!\!\sum\limits_{m=1}^M  \left[\left(\gamma_{l_m}^{({\text c})}\right)^2 {\text {MSE}}_{l_m}^{({\text c})} - 2 \ln \gamma_{l_m}^{({\text c})} -1 \right] \nonumber\\
= &&\!\!\sum\limits_{m=1}^M \left\{ \left(\gamma_{l_m}^{({\text c})}\right)^2 {\mathbb E} \left\{ \left|{\bm w}_{l_m}^H {\bm y}_l^{({\text c})}- {x}_{l_m}^{({\text c})} \right|^2\right\}- 2\ln \gamma_{l_m}^{({\text c})} -1 \right\}\nonumber\\
= &&\!\! \sum\limits_{m=1}^M\left\|q_{l_m}^{({\text c})} {\bm J}_l^H {\bm h}_{l_m,l}^{({\text c})} - \gamma_{l_m}^{({\text c})} {\bm e}_m\right\|^2 \nonumber\\
&&\!+ \sum\limits_{k_t \in {\cal C}\setminus{\cal C}_l} \left\|q_{k_t}^{({\text c})} {\bm J}_l^H {\bm h}_{k_t,l}^{({\text c})} \right\|^2 + \sum\limits_{k_s \in {\cal D}} \left\|q_{k_s}^{({\text d})} {\bm J}_l^H {\bm h}_{k_s,l}^{({\text d})} \right\|^2 \nonumber\\
&& + \sum\limits_{m=1}^M N_0 \left\| {\bm J}_{l,m} \right\|^2 - 2\sum\limits_{m=1}^M \ln \gamma_{l_m}^{({\text c})} - M,
\label{MSE_CU}
\end{eqnarray}}
\!\!where ${\bm J}_l = {\bm W}_l {\bm \Gamma}_l$ and ${\bm J}_{l,m}$ denotes the $m$th column of ${\bm J}_l$, i.e., ${\bm J}_{l,m} = \gamma_{l_m}^{({\text c})} {\bm w}_{l_m}$. ${\bm e}_m$ is an $M$ dimensional vector with one in the $m$th position and zeros elsewhere. By introducing a slack scalar variable $r_l$ to bound (\ref{MSE_CU}), (\ref{optimize_problem6}) can be reformulated in the following form
{\setlength\arraycolsep{2pt}
\begin{subequations}
\begin{align}
\mathop {\min }\limits_{\bm J_l, \bm \Gamma_l, r_l} \quad& r_l  \label{optimize_problem7_a}\\
\text{s.t.} \quad\quad & \sum\limits_{m=1}^M\left\|q_{l_m}^{({\text c})} {\bm J}_l^H {\bm h}_{l_m,l}^{({\text c})} - \gamma_{l_m}^{({\text c})} {\bm e}_m\right\|^2 \nonumber\\
& + \sum\limits_{k_t \in {\cal C}\setminus{\cal C}_l} \left\|q_{k_t}^{({\text c})} {\bm J}_l^H {\bm h}_{k_t,l}^{({\text c})} \right\|^2 + \sum\limits_{k_s \in {\cal D}} \left\|q_{k_s}^{({\text d})} {\bm J}_l^H {\bm h}_{k_s,l}^{({\text d})} \right\|^2 \nonumber\\
& + \!\sum\limits_{m=1}^M\! N_0 \left\| {\bm J}_{l,m} \right\|^2 - 2\sum\limits_{m=1}^M\! \ln \gamma_{l_m}^{({\text c})} \!-\! M \leq r_l,\label{optimize_problem7_b}\\
&\left\| {\bm \Delta}_{k_t,l}^{({\text c})} \right\| \leq \varepsilon_{k_t,l}^{({\text c})}, ~\forall~ k_t\in {\cal C},\label{optimize_problem7_c}\\
&\left\| {\bm \Delta}_{k_s,l}^{({\text d})} \right\| \leq \varepsilon_{k_s,l}^{({\text d})}, ~\forall~ k_s \in {\cal D},\label{optimize_problem7_d}\\
&\gamma_{l_m}^{({\text c})} > 0, ~\forall~ l_m \in {\cal C}_l.\label{optimize_problem7_e}
\end{align}
\label{optimize_problem7}
\end{subequations}}
\!\!\!\!\!\!\!\!To further simplify constraint (\ref{optimize_problem7_b}), additional auxiliary variables ${\bm b}_l = \left( b_{1_1,l}^{({\text c})}, \cdots, b_{1_M,l}^{({\text c})}, \cdots,b_{L_M,l}^{({\text c})}, b_{1_1,l}^{({\text d})}, \cdots, b_{1_N,l}^{({\text d})},\right.$ $\left. \cdots,b_{L_N,l}^{({\text d})}\right)^T$, ${\bm o}_l=(o_{l_1}, \cdots, o_{l_M})^T$ are introduced, and (\ref{optimize_problem7_b}) with (\ref{optimize_problem7_c}), (\ref{optimize_problem7_d}) are rewritten as follows
{\setlength\arraycolsep{2pt}
\begin{subequations}
\begin{align}
\!\!\!& \left\|q_{l_m}^{({\text c})} {\bm J}_l^H {\bm h}_{l_m,l}^{({\text c})} - \gamma_{l_m}^{({\text c})} {\bm e}_m\right\| \leq b_{l_m,l}^{({\text c})}, \nonumber\\
\!\!\!& \quad \forall~ \left\| {\bm \Delta}_{l_m,l}^{({\text c})} \right\| \leq \varepsilon_{l_m,l}^{({\text c})},~ \forall~ l_m\in {\cal C}_l, \label{constraints1_a}\\
\!\!\!& \left\|q_{k_t}^{({\text c})} {\bm J}_l^H {\bm h}_{k_t,l}^{({\text c})} \right\| \leq b_{k_t,l}^{({\text c})}, ~\forall~ \left\| {\bm \Delta}_{k_t,l}^{({\text c})} \right\| \leq \varepsilon_{k_t,l}^{({\text c})}, ~\forall~ k_t \in {\cal C}\setminus {\cal C}_l,\label{constraints1_b}\\
\!\!\!& \left\|q_{k_s}^{({\text d})} {\bm J}_l^H {\bm h}_{k_s,l}^{({\text d})} \right\| \leq b_{k_s,l}^{({\text d})}, ~\forall~ \left\| {\bm \Delta}_{k_s,l}^{({\text d})} \right\| \leq \varepsilon_{k_s,l}^{({\text d})}, ~\forall~ k_s\in {\cal D},\label{constraints1_c}\\
\!\!\!& \sqrt{N_0}\left\| {\bm J}_{l,m} \right\| \leq o_{l_m}, ~\forall~ l_m \in  {\cal C}_l,\label{constraints1_d}\\
\!\!\!& \sum\limits_{k_t \in {\cal C}} \left( b_{k_t,l}^{({\text c})}\right)^2 + \sum\limits_{k_s \in {\cal D}} \left( b_{k_s,l}^{({\text d})}\right)^2 + \sum\limits_{m=1}^M o_{l_m}^2 \nonumber\\
\!\!\!& - 2\sum\limits_{m=1}^M \ln \gamma_{l_m}^{({\text c})} - M \leq r_l.\label{constraints1_e}
\end{align}
\label{constraints1}
\end{subequations}}
\!\!\!\!Obviously, (\ref{constraints1_d}) is a second-order cone (SOC) constraint and (\ref{constraints1_e}) is a convex constraint. As for constraints (\ref{constraints1_a})$\sim$(\ref{constraints1_c}), they can be transformed into finite linear matrix inequalities (LMIs). First take (\ref{constraints1_a}) for instance. By applying the Schur Complement Lemma \cite{horn2012matrix}, constraint (\ref{constraints1_a}) can be equivalently stated as
{\setlength\arraycolsep{2pt}
\begin{eqnarray}
\!\!\!\!\!&& \left[ \begin{array}{l} \quad\quad\quad\quad b_{l_m,l}^{({\text c})}  \quad\quad\quad\;\;q_{l_m}^{({\text c})} \left(\!{\tilde {\bm h}}_{l_m,l}^{({\text c})} \!+\! {\bm \Delta}_{l_m,l}^{({\text c})}\!\right)^H \!\!{\bm J}_l \!-\! \gamma_{l_m}^{({\text c})} {\bm e}_m^H\\
\vspace{0.01em}\\
\!\!q_{l_m}^{({\text c})} {\bm J}_l^H \!\left(\!{\tilde {\bm h}}_{l_m,l}^{({\text c})} \!+\! {\bm \Delta}_{l_m,l}^{({\text c})}\!\right)\!-\! \gamma_{l_m}^{({\text c})} {\bm e}_m\quad\quad\quad b_{l_m,l}^{({\text c})} {\bm I} \end{array} \!\right] \nonumber\\
\!\!\!\!\!&& \quad\succeq 0, ~\forall~ \left\| {\bm \Delta}_{l_m,l}^{({\text c})} \right\| \leq \varepsilon_{l_m,l}^{({\text c})}, ~\forall~ l_m\in {\cal C}_l,
\label{constraints2}
\end{eqnarray}}
\!\!Denote ${\bm X}=[-1 \;\;{\bm 0}]$, ${\bm Y}=\left[{\bm 0}\;\; q_{l_m}^{({\text c})} {\bm J}_l\right]$, ${\bm Z}=\left( {\bm \Delta}_{l_m,l}^{({\text c})} \right)^H$ and
\begin{equation}
{\bm \Lambda}= \left[\! \begin{array}{l} \quad\quad b_{l_m,l}^{({\text c})}  \quad\quad\quad\;\;q_{l_m}^{({\text c})} \left({\tilde {\bm h}}_{l_m,l}^{({\text c})}\right)^H \!\!{\bm J}_l \!-\! \gamma_{l_m}^{({\text c})} {\bm e}_m^H\\
\vspace{0.01em}\\
\!\!q_{l_m}^{({\text c})} {\bm J}_l^H {\tilde {\bm h}}_{l_m,l}^{({\text c})}\!-\! \gamma_{l_m}^{({\text c})} {\bm e}_m\quad\quad\quad b_{l_m,l}^{({\text c})} {\bm I} \end{array} \!\!\right]\!.
\end{equation}
Then, based on Lemma \ref{lemma_1}, constraint (\ref{constraints1_a}) can be equivalently represented by
{\setlength\arraycolsep{2pt}
\begin{eqnarray}
\!\!\!\!\!&& {\bm F}_{l_m,l}^{({\text c})} = \left[ \begin{array}{l} \quad b_{l_m,l}^{({\text c})}\!-\! \eta_{l_m,l}^{({\text c})} \quad q_{l_m}^{({\text c})}\left(\!{\tilde {\bm h}}_{l_m,l}^{({\text c})}\!\right)^H \!\!\!{\bm J}_l \!-\! \gamma_{l_m}^{({\text c})} {\bm e}_m^H \quad \quad{\bm 0}\\
\vspace{0.01em}\\
\!\!q_{l_m}^{({\text c})} {\bm J}_l^H {\tilde {\bm h}}_{l_m,l}^{({\text c})}\!-\! \gamma_{l_m}^{({\text c})} {\bm e}_m\quad\quad b_{l_m,l}^{({\text c})} {\bm I} \quad\; -\varepsilon_{l_m,l}^{({\text c})} q_{l_m}^{({\text c})} {\bm J}_l^H\\
\vspace{0.01em}\\
\quad\quad\quad {\bm 0} \quad\quad\quad\quad\quad\; -\varepsilon_{l_m,l}^{({\text c})} q_{l_m}^{({\text c})} {\bm J}_l \quad\quad \eta_{l_m,l}^{({\text c})}{\bm I}\end{array} \!\!\right]\!\! \nonumber\\
\!\!\!\!\!&& \quad\succeq 0, ~\forall~ l_m\in {\cal C}_l,
\label{constraints3}
\end{eqnarray}}
\!\!which is a LMI and can be surely satisfied by finding a proper $\eta_{l_m,l}^{({\text c})}\geq 0$. Similarly, constraints (\ref{constraints1_b}) and (\ref{constraints1_c}) can also be transformed into LMIs. Thus, problem (\ref{optimize_problem6}) can be cast to an SDP as follows
{\setlength\arraycolsep{2pt}
\begin{subequations}
\begin{align}
\!\!\!\mathop {\min }\limits_{\bm J_l, \bm \Gamma_l, r_l, \bm b_l, \bm o_l, \bm \eta_l} \;& r_l  \label{optimize_problem10_a}\\
\text{s.t.} \quad\quad & {\bm F}_{k_t,l}^{({\text c})}\succeq 0, ~\forall~ k_t\in {\cal C} ,\label{optimize_problem10_b}\\
\!\!\!& {\bm F}_{k_s,l}^{({\text d})}\succeq 0, ~\forall~ k_s\in {\cal D} ,\label{optimize_problem10_c}\\
\!\!\!& \eta_{k_t,l}^{({\text c})} \geq 0, ~\forall~ k_t\in {\cal C},\label{optimize_problem10_d}\\
\!\!\!& \eta_{k_s,l}^{({\text d})} \geq 0, ~\forall~ k_s \in {\cal D},\label{optimize_problem10_e}\\
\!\!\!& {\text {(\ref{constraints1_d})}},~ {\text {(\ref{constraints1_e})}},~ {\text {(\ref{optimize_problem7_e})}},
\end{align}
\label{optimize_problem10}
\end{subequations}}
\!\!\!where ${\bm \eta}_l \!=\! ( \eta_{1_1,l}^{({\text c})}, \cdots, \eta_{1_M,l}^{({\text c})}, \cdots,\eta_{L_M,l}^{({\text c})}, \eta_{1_1,l}^{({\text d})}, \cdots, \eta_{1_N,l}^{({\text d})}, \cdots,$ $\eta_{L_N,l}^{({\text d})})^T$, ${\bm F}_{k_t,l}^{({\text c})}, ~\forall~ k_t\in {\cal C}\setminus {\cal C}_l$ and ${\bm F}_{k_s,l}^{({\text d})}$ are given by
{\setlength\arraycolsep{2pt}
\begin{eqnarray}
\!\!\!\!\!&& {\bm F}_{k_t,l}^{({\text c})} = \left[ \begin{array}{l} b_{k_t,l}^{({\text c})}\!-\! \eta_{k_t,l}^{({\text c})} \quad q_{k_t}^{({\text c})} \left(\!{\tilde {\bm h}}_{k_t,l}^{({\text c})}\!\right)^H \!\!\!{\bm J}_l \quad \quad{\bm 0}\\
\vspace{0.01em}\\
q_{k_t}^{({\text c})} {\bm J}_l^H {\tilde {\bm h}}_{k_t,l}^{({\text c})}\quad\quad b_{k_t,l}^{({\text c})} {\bm I} \quad -\varepsilon_{k_t,l}^{({\text c})} q_{k_t}^{({\text c})} {\bm J}_l^H\\
\vspace{0.01em}\\
\quad\quad {\bm 0} \quad\quad\quad -\varepsilon_{k_t,l}^{({\text c})} q_{k_t}^{({\text c})} {\bm J}_l \;\;\quad \eta_{k_t,l}^{({\text c})}{\bm I}\end{array} \right], \nonumber\\
\!\!\!\!\!&& \quad \forall~ k_t \in  {\cal C}\setminus{\cal C}_l,
\end{eqnarray}}
\!\!\!and
\begin{equation}
\!\!{\bm F}_{k_s,l}^{({\text d})} \!=\!\! \left[\!\!\! \begin{array}{l} b_{k_s,l}^{({\text d})}\!-\! \eta_{k_s,l}^{({\text d})} \quad q_{k_s}^{({\text d})} \left(\!{\tilde {\bm h}}_{k_s,l}^{({\text d})}\!\right)^H \!\!\!{\bm J}_l \quad \quad{\bm 0}\\
\vspace{0.01em}\\
q_{k_s}^{({\text d})} {\bm J}_l^H {\tilde {\bm h}}_{k_s,l}^{({\text d})}\quad\quad b_{k_s,l}^{({\text d})} {\bm I} \quad -\varepsilon_{k_s,l}^{({\text d})} q_{k_s}^{({\text d})} {\bm J}_l^H\\
\vspace{0.01em}\\
\quad\quad {\bm 0} \quad\quad\quad -\varepsilon_{k_s,l}^{({\text d})} q_{k_s}^{({\text d})} {\bm J}_l \;\;\quad \eta_{k_s,l}^{({\text d})}{\bm I}\end{array} \!\!\!\!\right]\!\!, ~\forall~ k_s\in {\cal D}.
\end{equation}
Thus far, problem (\ref{optimize_problem10}) has been recognized as an SDP with linear objective and LMI or SOC constraints. Once $\bm J_l$ and $\bm \Gamma_l$ have been obtained by solving (\ref{optimize_problem10}), one can readily get $\bm W_l$ based on the relationship $\bm W_l=\bm J_l \bm \Gamma_l^{-1}$.

\section{Proof of Theorem \ref{theorem_2}}
\label{Appendix_C}
\subsection{Casting (\ref{optimize_problem8}) to an SDP}
\label{Appendix_C_A}
Similar as the proof in Appendix B, (\ref{optimize_problem8}) can be transformed to the following form by introducing auxiliary variables $\beta_{l_m}$ and $\bm \xi_{l_m}^{({\text c})}=(\xi_{l_m,1}^{({\text c})},\cdots,\xi_{l_m,L}^{({\text c})})^T$
\setcounter{equation}{0}
{\setlength\arraycolsep{2pt}
\begin{subequations}
\begin{align}
\!\!\!\mathop {\min }\limits_{q_{l_m}^{({\text c})}, \beta_{l_m}, \bm \xi_{l_m}^{({\text c})}} \;& \beta_{l_m}  \label{optimize_problem15_a}\\
\text{s.t.} \quad\quad & \left\| q_{l_m}^{({\text c})} {\bm J}_l^H {\bm h}_{l_m,l}^{({\text c})} - \gamma_{l_m}^{({\text c})} {\bm e}_m\right\| \leq \xi_{l_m,l}^{({\text c})}, \label{optimize_problem15_b}\\
\!\!\!& \left\| q_{l_m}^{({\text c})} {\bm J}_k^H {\bm h}_{l_m,k}^{({\text c})} \right\| \leq \xi_{l_m,k}^{({\text c})}, ~\forall~ k\in {\cal L}\setminus l,\label{optimize_problem15_c}\\
\!\!\!& \sum\limits_{k \in {\cal L}} \left(\xi_{l_m,k}^{({\text c})}\right)^2 + \left(q_{l_m}^{({\text c})}\right)^2 \sum\limits_{k_s \in {\cal D}} \left|\gamma_{k_s}^{({\text d})} f_{k_s}^H {\bar g}_{l_m,k_s}^{({\text c})} \right|^2 \nonumber\\
\!\!\!&\leq \beta_{l_m}^2,\label{optimize_problem15_d}\\
\!\!\!& \beta_{l_m}\geq 0, \label{optimize_problem15_e}\\
& 0 \leq q_{l_m}^{({\text c})} \leq \sqrt{P_{l_m}^{({\text c})}},\label{optimize_problem15_f}\\
&\left\| {\bm \Delta}_{l_m,k}^{({\text c})} \right\| \leq \varepsilon_{l_m,k}^{({\text c})}, ~\forall~ k\in {\cal L}.\label{optimize_problem15_g}
\end{align}
\label{optimize_problem15}
\end{subequations}}
\!\!\!\!\!\!\!Applying the Schur Complement Lemma and Lemma \ref{lemma_1}, constraints (\ref{optimize_problem15_b}), (\ref{optimize_problem15_c}) and (\ref{optimize_problem15_g}) can be transformed into LMIs by further introducing auxiliary variable ${\bm \varphi}_{l_m}^{({\text c})} = ( \varphi_{l_m,1}^{({\text c})}, \cdots, \varphi_{l_m,L}^{({\text c})})^T$. Accordingly, problem (\ref{optimize_problem15}) can be rewritten as
{\setlength\arraycolsep{2pt}
\begin{subequations}
\begin{align}
\!\!\!\mathop {\min }\limits_{q_{l_m}^{({\text c})}, \beta_{l_m}, \bm \xi_{l_m}^{({\text c})}, {\bm \varphi}_{l_m}^{({\text c})}} \;& \beta_{l_m}  \label{optimize_problem16_a}\\
\text{s.t.} \quad\quad & {\bm U}_{l_m,k}^{({\text c})}\succeq 0, ~\forall~ k\in {\cal L} ,\label{optimize_problem16_b}\\
\!\!\!& \varphi_{l_m,k}^{({\text c})} \geq 0, ~\forall~ k\in {\cal L},\label{optimize_problem16_c}\\
\!\!\!& {\text {(\ref{optimize_problem15_d})}}\sim {\text {(\ref{optimize_problem15_f})}},\label{optimize_problem16_d}
\end{align}
\label{optimize_problem16}
\end{subequations}}
\!\!\!\!where ${\bm U}_{l_m,k}$ is given by
\begin{equation}
\!{\bm U}_{l_m,l}^{({\text c})} \!=\!\! \left[\! \begin{array}{l} \; \xi_{l_m,l}^{({\text c})}\!-\! \varphi_{l_m,l}^{({\text c})} \quad\; q_{l_m}^{({\text c})}\! \left(\!{\tilde {\bm h}}_{l_m,l}^{({\text c})}\!\right)^H \!\!\!{\bm J}_l \!-\! \gamma_{l_m}^{({\text c})} {\bm e}_m^H \quad \quad{\bm 0}\\
\vspace{0.05em}\\
\!\!q_{l_m}^{({\text c})} {\bm J}_l^H {\tilde {\bm h}}_{l_m,l}^{({\text c})}\!-\! \gamma_{l_m}^{({\text c})} {\bm e}_m\quad\quad \xi_{l_m,l}^{({\text c})} {\bm I} \quad\; -\varepsilon_{l_m,l}^{({\text c})} q_{l_m}^{({\text c})} {\bm J}_l^H\\
\vspace{0.05em}\\
\quad\quad\; {\bm 0} \quad\quad\quad\quad\quad -\varepsilon_{l_m,l}^{({\text c})} q_{l_m}^{({\text c})} {\bm J}_l \quad\quad\quad\; \varphi_{l_m,l}^{({\text c})} {\bm I}\end{array} \!\!\!\!\right]\!\!,\!
\end{equation}
and
{\setlength\arraycolsep{2pt}
\begin{eqnarray}
\!\!\!\!\!&& {\bm U}_{l_m,k}^{({\text c})} = \left[ \begin{array}{l}  \xi_{l_m,k}^{({\text c})}\!-\! \varphi_{l_m,k}^{({\text c})} \quad\! q_{l_m}^{({\text c})}\! \left(\!{\tilde {\bm h}}_{l_m,k}^{({\text c})}\!\right)^H \!\!\!{\bm J}_k  \quad \quad \! {\bm 0}\\
\vspace{0.01em}\\
q_{l_m}^{({\text c})} {\bm J}_k^H {\tilde {\bm h}}_{l_m,k}^{({\text c})}\quad\quad\!\! \xi_{l_m,k}^{({\text c})} {\bm I} \quad\! -\varepsilon_{l_m,k}^{({\text c})} q_{l_m}^{({\text c})} {\bm J}_k^H\\
\vspace{0.01em}\\
\quad\quad {\bm 0} \quad\quad\;\;\;\;\! -\varepsilon_{l_m,k}^{({\text c})} q_{l_m}^{({\text c})} {\bm J}_k \quad\;\! \varphi_{l_m,k}^{({\text c})} {\bm I}\end{array} \right], \nonumber\\
\!\!\!\!\!&& \quad \forall~ k \!\in\! {\cal L}\setminus l.
\end{eqnarray}}
\!\!\!Obviously, problem (\ref{optimize_problem16}) is an SDP.
\subsection{Casting (\ref{optimize_problem5}) as an SDP}
\label{Appendix_C_B}
Analogous to the procedure in the above subsection, problem (\ref{optimize_problem5}) can be cast to an SDP as follows
{\setlength\arraycolsep{2pt}
\begin{subequations}
\begin{align}
\!\!\!\mathop {\min }\limits_{\bm q^{({\text d})}, v, \bm \xi^{({\text d})}, \bm \varphi^{({\text d})}} \;& v  \label{optimize_problem21_a}\\
\text{s.t.} \quad\; & {\bm U}_{l_n,k}^{({\text d})}\succeq 0, ~\forall~ l_n\in {\cal D},~ k \in {\cal L},\label{optimize_problem21_b}\\
\!\!\!& \varphi_{l_n,k}^{({\text d})} \geq 0, ~\forall~ l_n\in {\cal D},~ k \in {\cal L},\label{optimize_problem21_c}\\
\!\!\!& \sum\limits_{l_n \in {\cal D}} \!\left\{\!\sum\limits_{k=1}^L \left(\xi_{l_n,k}^{({\text d})}\right)^2 \!+\! \theta_{l_n}\! \left(\!q_{l_n}^{({\text d})}\!\right)^2 \!-\! 2 \phi_{l_n} q_{l_n}^{({\text d})}\!\right\}\nonumber\\
\!\!\!& \leq v,\label{optimize_problem21_d}\\
\!\!\!& {\text {(\ref{optimize_problem5_b})}}, {\text {(\ref{optimize_problem5_c})}}, \label{optimize_problem21_e}
\end{align}
\label{optimize_problem21}
\end{subequations}}
\!\!\!\!\!\!\!where $v$, ${\bm \xi}^{({\text d})} = ( \xi_{1_1,1}^{({\text d})}, \cdots, \xi_{1_1,L}^{({\text d})}, \cdots,\xi_{L_N,L}^{({\text d})})^T$, ${\bm \varphi}^{({\text d})}$ $ =( \varphi_{1_1,1}^{({\text d})}, \cdots, \varphi_{1_1,L}^{({\text d})}, \cdots,\varphi_{L_N,L}^{({\text d})})^T$ are auxiliary variables, and ${\bm U}_{l_n,k}^{({\text d})}$ is given by
{\setlength\arraycolsep{2pt}
\begin{eqnarray}
&&{\bm U}_{l_n,k}^{({\text d})} = \left[ \begin{array}{l}  \xi_{l_n,k}^{({\text d})}\!-\! \varphi_{l_n,k}^{({\text d})} \quad q_{l_n}^{({\text d})}\! \left(\!{\tilde {\bm h}}_{l_n,k}^{({\text d})}\!\right)^H \!\!\!{\bm J}_k  \quad \quad\;\; {\bm 0}\\
\vspace{0.01em}\\
\!\!q_{l_n}^{({\text d})} {\bm J}_k^H {\tilde {\bm h}}_{l_n,k}^{({\text d})}\quad\quad \xi_{l_n,k}^{({\text d})} {\bm I} \quad\quad -\varepsilon_{l_n,k}^{({\text d})} q_{l_n}^{({\text d})} {\bm J}_k^H\\
\vspace{0.01em}\\
\quad\quad {\bm 0} \quad\quad\quad\; -\varepsilon_{l_n,k}^{({\text d})} q_{l_n}^{({\text d})} {\bm J}_k \quad\quad \varphi_{l_n,k}^{({\text d})}{\bm I}\end{array} \!\!\right], \nonumber\\
\quad &&\quad \forall~ l_n\in {\cal D},~ k \in {\cal L}.
\end{eqnarray}}

Then, Theorem \ref{theorem_2} is proven.
\section{Proof of Theorem \ref{theorem_3}}
\label{Appendix_D}
For notational convenience, denote the objective function of (\ref{optimize_problem3}) by
\setcounter{equation}{0}
{\setlength\arraycolsep{2pt}
\begin{eqnarray}
&&V({\bm p},{\bm W},{\bm f},{\bm u})\nonumber\\
&=& \sum\limits_{l=1}^L \mathop {\max }\limits_{{\bm \Delta}_l} \sum\limits_{m=1}^M \left[ \exp \left(u_{l_m}^{({\text c})}-1\right){\text {MSE}}_{l_m}^{({\text c})} - u_{l_m}^{({\text c})} \right]\nonumber\\
&& + \sum\limits_{l_n \in {\cal D}} \left[ \exp \left(u_{l_n}^{({\text d})} - 1\right) {\text {MSE}}_{l_n}^{({\text d})} - u_{l_n}^{({\text d})} \right].
\label{objective_func}
\end{eqnarray}}
\!\!To verify the convergence of Algorithm \ref{algorithm1}, it is first shown that (\ref{objective_func}) is lower bounded. By dropping the positive interference terms in the denominator of ${\text {SINR}}_{l_m}^{({\text c})}$ and using the Cauchy-Schwartz inequality, the SINR of CU $l_m$ is upper bounded by
{\setlength\arraycolsep{2pt}
\begin{eqnarray}
{\text {SINR}}_{l_m}^{({\text c})} &\leq& \frac{p_{l_m}^{({\text c})} \left|\left(\!{\bm h}_{l_m,l}^{({\text c})}\right)^H {\bm w}_{l_m} \right|^2}{N_0 \left\|{\bm w}_{l_m}\right\|^2}\leq \frac{p_{l_m}^{({\text c})} \left\|{\bm h}_{l_m,l}^{({\text c})}\right\|^2 \left\| {\bm w}_{l_m} \right\|^2}{N_0 \left\|{\bm w}_{l_m}\right\|^2}\nonumber\\
&\leq& \frac{P_{l_m}^{({\text c})}}{N_0} \left(\left\|{\tilde {\bm h}}_{l_m,l}^{({\text c})}\right\| + \varepsilon_{l_m,l}^{({\text c})}\right)^2 \triangleq \sigma_{l_m}^{({\text c})}, ~\forall~ l_m\in {\cal C}.\quad\quad
\label{SINR_upper_CU}
\end{eqnarray}}
\!\!\!Similarly, the SINR of D2D link $l_n$ is upper bounded by
\begin{equation}
{\text {SINR}}_{l_n}^{({\text d})} \leq \frac{ P_{l_n}^{({\text d})}}{N_0} \left|g_{l_n,l_n}^{({\text d})}\right|^2 \triangleq \sigma_{l_n}^{({\text d})}, ~\forall~ l_n\in {\cal D}.
\label{SINR_upper_DU}
\end{equation}
Hence, a lower bound on the objective function of (\ref{optimize_problem26}) is
\begin{equation}
\sum\limits_{l_m \in {\cal C}}  \ln \frac{1}{1+\sigma_{l_m}^{({\text c})}} + \sum\limits_{l_n \in {\cal D}} \ln \frac{1}{1+\sigma_{l_n}^{({\text d})}}.
\label{object_lower_bound}
\end{equation}
As mentioned in Subsection \ref{Problem_Analysis}, the objective function of (\ref{optimize_problem26}) is upper bounded by (\ref{objective_func}). Therefore, (\ref{object_lower_bound}) is also a lower bound to (\ref{objective_func}).

Next, it is shown that the objective function of (\ref{optimize_problem3}) is non-increasing in each iteration. Without loss of generality, denote the solution obtained in the $j$th iteration by $\left\{{\bm p}(j), {\bm W}(j), {\bm f}(j), {\bm u}(j)\right\}$. Then, in the $(j+1)$th iteration, Algorithm \ref{algorithm1} starts by solving (\ref{optimize_problem10}), which is an SDP problem and can be optimally solved by applying CVX. So the optimal $\bm W(j+1)$ and $\bm u^{({\text c})}(j+1)$ can be obtained according to the relationships $\bm W_l(j+1) = \bm J_l(j+1)\bm \Gamma_l(j+1)^{-1}$ and $u_{l_m}^{({\text c})}(j+1)=1+2\ln \gamma_{l_m}^{({\text c})}(j+1), ~\forall~ l \in {\cal L},~ l_m \in {\cal C}_l$. Moreover, the optimal $f_{l_n}^{({\text d})}(j+1)$ and $u_{l_n}^{({\text d})}(j+1), ~\forall~ l_n \in {\cal D}$ can be calculated based on (\ref{MMSE_filter_DU}) and (\ref{auxiliary_DU3}). This yields the following relationship
{\setlength\arraycolsep{2pt}
\begin{eqnarray}
&&V\left({\bm p}(j),{\bm W}(j),{\bm f}(j),{\bm u}(j)\right)\nonumber\\
 &&\geq V\left({\bm p}(j),{\bm W}(j+1),{\bm f}(j+1),{\bm u}(j+1)\right).
\label{relationship1}
\end{eqnarray}}
\!\!For given ${\bm W}(j+1)$, ${\bm f}(j+1)$ and ${\bm u}(j+1)$, the optimal ${\bm p}(j+1)$ can be obtained by solving SDPs (\ref{optimize_problem16}) and (\ref{optimize_problem21}), and using the relationships $p_{l_m}^{({\text c})}(j+1)=\left[q_{l_m}^{({\text c})}(j+1)\right]^2, ~\forall~ l_m \in {\cal C}$, $p_{l_n}^{({\text d})}(j+1)=\left[q_{l_n}^{({\text d})}(j+1)\right]^2, ~\forall~ l_n \in {\cal D}$. Then,
{\setlength\arraycolsep{2pt}
\begin{eqnarray}
&&V\left({\bm p}(j),{\bm W}(j+1),{\bm f}(j+1),{\bm u}(j+1)\right)\nonumber\\
 &&\geq V\left({\bm p}(j+1),{\bm W}(j+1),{\bm f}(j+1),{\bm u}(j+1)\right).
\label{relationship2}
\end{eqnarray}}
\!\!Incorporating (\ref{relationship1}) and (\ref{relationship2}) yields
{\setlength\arraycolsep{2pt}
\begin{eqnarray}
&&V\left({\bm p}(j),{\bm W}(j),{\bm f}(j),{\bm u}(j)\right)\nonumber\\
 &&\geq V\left({\bm p}(j+1),{\bm W}(j+1),{\bm f}(j+1),{\bm u}(j+1)\right),
\label{relationship3}
\end{eqnarray}}
\!\!\!which indicates that the objective function of (\ref{optimize_problem3}) is non-increasing in each iteration. Noting the fact that (\ref{objective_func}) is lower bounded by (\ref{object_lower_bound}), it can be concluded that Algorithm \ref{algorithm1} converges to a suboptimal solution of problem (\ref{optimize_problem3}).

\bibliographystyle{IEEEtran}
\bibliography{IEEEabrv,MMM}

\end{document}